\documentclass{ws-rv975x65}
\usepackage[square]{ws-rv-van}
\makeindex

\def\m@th{\mathsurround=0pt }
\def\leftrightarrowfill{$\m@th \mathord\leftarrow \mkern-6mu
	\cleaders\hbox{$\mkern-2mu \mathord- \mkern-2mu$}\hfill
	\mkern-6mu \mathord\rightarrow$}

\def\overleftrightarrow#1{\vbox{\ialign{##\crcr
	\leftrightarrowfill\crcr\noalign{\kern-1pt\nointerlineskip}
	$\hfil\displaystyle{#1}\hfil$\crcr}}}

\newcommand{\be}{\begin{equation}}
\newcommand{\ee}{\end{equation}}

\newcommand{\Tr}{\mathop{\rm Tr}}
\def\I{\rm 1\kern-.24em l}  % Yes, I know. This ain't capital.

\def\shat{\ifmmode \hat{s}\else $\hat{s}$\fi}

\def\UV{UV}
\def\IR{IR}
\def\ZZ{\mathbb Z}
\def\nn{\nonumber}

\newcommand{\newc}{\newcommand}

\newc{\gsim}{\lower.7ex\hbox{$\;\stackrel{\textstyle>}{\sim}\;$}}
\newc{\lsim}{\lower.7ex\hbox{$\;\stackrel{\textstyle<}{\sim}\;$}}
\newc{\ie}{{\it i.e.}}
\newc{\etal}{{\it et al.}}
\newc{\mev}{\hbox{\rm\,MeV}}
\newc{\gev}{\hbox{\rm\,GeV}}
\newc{\tev}{\hbox{\rm\,TeV}}
\newc{\xpb}{\hbox{\rm\, pb}}
\newc{\xfb}{\hbox{\rm\, fb}}

\newc{\G}{{\cal G}}
\newc{\h}{{\cal H}}
\newc{\D}{{\cal D}}
\newc{\E}{{\cal E}}
\newc{\x}{{\widehat x}}
\newc{\q}{{\widehat q}}

\newc{\mtop}{m_t}
\newc{\mbot}{m_b}
\newc{\mz}{M_Z}
\newc{\mw}{M_W}
\newc{\alphasmz}{\alpha_s(M_Z)}
\newc{\swsq}{\sin^2\theta_W}
\newc{\cwsq}{\cos^2\theta_W}
\newc{\tw}{\tan\theta_W}
\newc{\cw}{\cos\theta_W}
\newc{\sw}{\sin\theta_W}
\newc{\BR}{\hbox{\rm BR}}
\newc{\zbb}{Z\to b\bar}
\newc{\Gb}{\Gamma (Z\to b\bar b)}
\newc{\Gh}{\Gamma (Z\to \hbox{\rm hadrons})}
\newc{\sgn}{\mbox{sgn}}

\def\I{1\hspace{-4pt}1}
\def\ov{\overline}

\newcounter{mysubequation}[equation]

\def\beq{\begin{equation}}
\def\eeq{\end{equation}}
\def\bea{\begin{eqnarray}}
\def\eea{\end{eqnarray}}
\def\slashchar#1{\setbox0=\hbox{$#1$}           % set a box for #1
   \dimen0=\wd0                                 % and get its size
   \setbox1=\hbox{/} \dimen1=\wd1               % get size of /
   \ifdim\dimen0>\dimen1                        % #1 is bigger
      \rlap{\hbox to \dimen0{\hfil/\hfil}}      % so center / in box
      #1                                        % and print #1
   \else                                        % / is bigger
      \rlap{\hbox to \dimen1{\hfil$#1$\hfil}}   % so center #1
      /                                         % and print /
   \fi}                                         %

\catcode`@=11
\long\def\@caption#1[#2]#3{\par\addcontentsline{\csname
  ext@#1\endcsname}{#1}{\protect\numberline{\csname
  the#1\endcsname}{\ignorespaces #2}}\begingroup
    \small
    \@parboxrestore
    \@makecaption{\csname fnum@#1\endcsname}{\ignorespaces #3}\par
  \endgroup}
\catcode`@=12

\def\UV{\rm UV}
\def\IR{\rm IR}
\def\Le{{\bf L}}
\def\R{{\bf R}}
\def\A{{\bf A}}
\def\J{{\bf J}}

\begin{document}

%%%%%%%%%%%%%%%%%%%%%%%%%%%%%

\chapter{Baryon physics in a five-dimensional model of hadrons
\label{ch1}}

\author[F. Author and S. Author]{Alex Pomarol\footnote{alex.pomarol@uab.cat}}
%{First Author and Second Author\footnote{Author footnote.}}
%\index[aindx]{Author, F.} % or \aindx{Author, F.}
%\index[aindx]{Author, S.} % or \aindx{Author, S.}

\address{Departament de F\'isica, Universitat Aut\`onoma de Barcelona,\\
08193 Bellaterra, Barcelona}
\author[F. Author and S. Author]{Andrea Wulzer\footnote{andrea.wulzer@epfl.ch}}
%\index[aindx]{Author, F.} % or \aindx{Author, F.}
%\index[aindx]{Author, S.} % or \aindx{Author, S.}
\address{Institut de Th\'eorie des Ph\'enom\`enes Physiques, EPFL,\\  CH--1015 Lausanne, Switzerland}

\begin{abstract}
We  review  the procedure to calculate  baryonic properties 
using   a  recently proposed five-dimensional approach   to  QCD. 
We show that this method give predictions to baryon observables  that  agree
reasonable well with   the experimental data.  
\end{abstract}

\body

%%%%%%%%%%%%%%%%%%%%%%%%%%
\section{Introduction}

In 1973 Gerard 'tHooft proposed, in a seminal article \cite{'tHooft:1973jz},
 a dual description for QCD.
He showed that in the limit of large number of colors ($N_c$)  strongly-interacting gauge  theories
 could be described  in terms of a weakly-interacting  
 theory  of  mesons. 
It was later recognized \cite{Adkins:1983ya} that,   in this dual description, baryons appeared
 as  solitons  made  of meson fields, as Skyrme had pointed out long before  \cite{Skyrme:1961vq}.
These solitonic states  were therefore   referred as skyrmions. 

Skyrmions have been widely studied in the literature, with some phenomenological successes
\cite{Meissner:1987ge}. 
Nevertheless, since the full theory of QCD mesons is not known, 
these studies  have been carried out in truncated low-energy models 
either  incorporating  only pions \cite{Skyrme:1961vq,Adkins:1983ya} or
 few  resonances  \cite{Meissner:1987ge}.
It is unclear whether  these approaches 
capture the physics  needed to fully describe the baryons, since
the stabilization of the baryon size is very sensitive to resonances around the GeV.
In the original Skyrme model with only pions, for instance, the inverse skyrmion size $\rho_s^{-1}$
equals the chiral perturbation theory cut-off $\Lambda_{\chi PT}\sim4\pi F_\pi$ (as it should be, since this is the only scale of the model),  rendering baryon physics completely  incalculable. 
Other examples are models with the $\rho$-meson \cite{Igarashi:1985et} or the $\omega$-meson
\cite{Adkins:1983nw} which were 
shown to  have a stable  
skyrmion solution. The inverse size, also in this case, is of order $m_\rho\sim\Lambda_{\chi PT}$, which is clearly not far from the mass of the next resonances.
Including the latter could affect strongly the physics of the skyrmion, or even destabilize it.

In the last ten years
the string/gauge duality  \cite{Maldacena:1997re,Gubser:1998bc,Witten:1998qj} has allowed us  to gain new insights into the
problem of strongly-coupled gauge theories. 
This duality  has been able
to relate certain strongly-coupled  gauge theories with  string theories living in more than four 
dimensions.
A crucial ingredient in these realizations is a (compact) warped extra dimension
that  plays the role of the energy scale in the strongly-coupled 4D theory. 
This has    suggested that  
the QCD dual  theory of mesons  proposed by 'tHooft \cite{'tHooft:1973jz} must be a 
theory formulated in more than 4 dimensions.

Inspired but this duality, a five-dimensional field theory  has been proposed 
in Refs.~\cite{Erlich:2005qh,Rold:2005} 
 to describe   the properties of mesons in QCD.  
This 5D theory
 has a  cut-off scale    $\Lambda_5$ 
which is   above the lowest-resonance mass $m_\rho$.
The gap among these two scales, which  ensures calculability in the meson sector, 
is related  to  the number of colors $N_c$ of QCD. In the large $N_c$-limit, 
one has $\Lambda_5/m_\rho\rightarrow\infty$ and  the 5D model describes a 
 theory of infinite mesonic resonances, corresponding to the Kaluza-Klein (KK) spectrum.
This 5D  model has provided a  quite accurate   description of   meson physics in terms 
of a very limited number of parameters.

\begin{figure}
\label{bola}
%\centerline{\psfig{file=skyrmion.pdf,width=5cm}}\includegraphics[width=0.46 \textwidth]{gESPlot}
\centerline{\includegraphics[width=0.36 \textwidth]{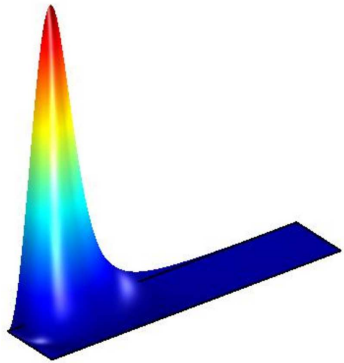}}
\caption{Energy density,
in the plane of the 4D radial and the extra fifth  coordinate,
of the skyrmion in a  5D model for QCD.}
\end{figure}

Further studies, boosted by this success,  have recently shown that the 5D model can
also successfully  describe  baryon physics 
\cite{Pomarol:2007kr,Pomarol:2008aa,Panico:2008it}. 
As Skyrme proposed \cite{Skyrme:1961vq},
baryons must appear in this 5D theory as   solitons.
These 5D skyrmion-like solitons  have been  numerically  obtained  
(see Fig.~\ref{bola})
and their properties have been  studied.
Their  inverse size $\rho_s^{-1}\sim m_\rho$ have been found to be smaller than the cut-off scale
$\Lambda_5$, showing then 
that, contrary to the 4D case, they can be consistently studied in 5D effective theories.
Indeed, the expansion parameter which ensures calculability is  provided by 
$1/(\rho_s\Lambda_5)\ll 1$. 

In this article we will review the properties of baryons obtained
in Refs.~\cite{Pomarol:2007kr,Pomarol:2008aa,Panico:2008it} using the  five-dimensional model   of QCD of Refs.~\cite{Erlich:2005qh,Rold:2005,Hirn:2005nr}.
We will show how the
calculation of the  static properties of the nucleons, 
such as masses, radii and form factors, 
 are performed, and   will compare the predictions of the model  with experiments.
As we will  see, these  predictions    
show a  reasonably   good agreement with the data.
 
There have been alternatives  studies to baryon physics using 5D models.
Nevertheless,  these studies  have encountered   several problems. 
For example, the first approaches \cite{Nawa:2006gv}
 truncated the 5D theory and only considered the effects of the first resonances.
This leads to skyrmions whose size
is of the order of the inverse of the truncation scale, and therefore sensitive to
the discarded heavier resonances.
Later studies  
\cite{Hata:2007mb,Hong:2006ta,Hata:2008xc} were performed within the 
Sakai-Sugimoto model \cite{Sakai:2004cn}. It was shown, however,     that baryons are not calculable  in this framework as their inverse size is of the order of the string scale which 
corresponds to the cut-off of the theory \cite{Hata:2007mb}.

%%%%%%%%%%%%%%%%%%%%%%%%%%%%%%%%%%%%%%%%%%%%%%%
\section{A five-dimensional  model for QCD  mesons}
\label{5Dmod}
%Description of the 5D   $U(2)^2$ model for hadrons.
%Sect.~2 of AA1(Alex-Andrea-1)

The 5D model that we will consider to describe mesons in two massless flavor QCD is the following.
This is  a $U(2)_L\times U(2)_R$ gauge theory with  metric  
$ds^2=a(z)^2\left(\eta_{\mu\nu}dx^\mu dx^\nu-dz^2\right)$, where $x^\mu$ represent the usual 
$4$ coordinates and $z$, which runs in the interval 
$[z_{\UV},z_{\IR}]$,  denotes the extra dimension. 
We will work in AdS$_5$ where the warp factor $a(z)$ is
\begin{equation}
a(z)=\frac{z_{\IR}}{z}\, ,
\label{warpf}
\end{equation}
and $z_{\UV}\rightarrow 0$ to  be taken at the end of the calculations.
In this limit $z_{\IR}$ coincides with the AdS curvature  and 
the conformal length
\be
  L=\int^{z_{\IR}}_{z_{\UV}} dz\,.
  \ee
The $U(2)_L$ and $U(2)_R$ gauge connections, denoted respectively by $\Le_M$ and 
$\R_M$ ($M=\{\mu,5\}$), are parametrized by $\Le_M=L_M^a\sigma_a/2+\widehat{L}_M\I/2$ and $\R_M=R_M^a\sigma_a/2+\widehat{R}_M\I/2$, where $\sigma_a$ are the Pauli matrices. 
This chiral gauge symmetry is broken by the conditions on the boundary at $z=z_{\IR}$ 
(IR-boundary), which read
\beq
\left(\Le_\mu-\R_\mu\right)\left|_{z=z_{\IR}}\right.=0\ ,\;\;\;\;\;\;\left(\Le_{\mu 5}+\R_{\mu 5}\right)\left|_{z=z_{\IR}}\right.=0\, ,
\label{irboundary condition}
\eeq
where the 5D field strength is defined as $\Le_{MN}=\partial_M \Le_N-\partial_N \Le_M-i[\Le_M,\,\Le_N]$, and analogously for $\R_{MN}$.
At the other boundary, the UV one, we can consider generalized Dirichlet conditions for all the fields:
\beq
 \Le_\mu\left|_{z=z_{\UV}}\right.=\,l_\mu\ , \;\;\;\;\;\; \R_\mu\left|_{z=z_{\UV}}\right.=\,r_\mu \, .
\label{uvboundary condition}
\eeq
The 4D fields $l_\mu$ and $r_\mu$ are arbitrary but fixed and they can be interpreted, as we will now 
discuss, as external sources for the QCD global currents. We will eventually be interested in taking 
the sources to vanish. 

We can now, inspired by the ``holographic" formulation 
 of the AdS/CFT correspondence \cite{Maldacena:1997re,Gubser:1998bc,Witten:1998qj},
try to interpret  the above  5D model in terms of a 4D 
QCD-like theory, whose fields we will generically denote by $\Psi(x)$ and its action by $S_4$. This is a  strongly coupled 4D theory that possesses an $U(2)_L\times U(2)_R$ global symmetry with  associated Noether currents  $j_{L,R}^\mu$. 
 If the 4D theory were precisely massless QCD with two flavors, the currents would be given by the usual quark bilinear, 
 $\left(j_{L,R}^\mu\right)_{ij}={\overline{Q}}_{L,R}^j\gamma^\mu Q_{L,R}^i$.  Defining $Z[l_\mu,r_\mu]$ as the generating functional of current correlators, we state our correspondence as
\bea
Z\left[l_\mu,r_\mu\right]\,\equiv&&\, 
\int \mathcal{D}\Psi\exp{\left[iS_4\left[\Psi\right]\,+\,i\int d^4 x {\textrm{Tr}}\left(j_{L}^\mu l_\mu+j_{R}^\mu r_\mu\right)\right]}\nn\\
\,=&&\,\int \mathcal{D}\Le_M\mathcal{D}\R_M\exp{\left[iS_5\left[\Le,\R\right]\right]}\, ,
\label{corr}
\eea
where the 5D partition function depends on the sources $l_\mu$, $r_\mu$ through 
the UV-boundary conditions in Eq.~(\ref{uvboundary condition}). 

Eq.~(\ref{corr}) leads to the following implication. Under local chiral transformations,
$Z$ receives a contribution from the $U(2)^3$ anomaly, which is known in QCD.
\footnote{
The 5D semiclassical expansion we perform in our model corresponds, as we will 
explain in the following, to the large-$N_c$ expansion.
This is why we are ignoring the $U(1)$-$SU(N_c)^2$ anomaly of QCD,
 which is subleading at large-$N_c$. Being this anomaly 
responsible for the $\eta^{\prime}$ mass, our model will contain a massless 
$\eta'$.}
This implies \cite{Witten:1998qj,Hill:2006wu,Panico:2007qd} that the 5D action must contain a Chern-Simons (CS) term
\be
S_{CS}\,=\,-i\frac{N_c}{24\pi^2}\int \left[\omega_5(\Le) - \omega_5(\R)\right]\, ,
\ee
whose variation under 5D local transformations which does not reduce to the identity at the 
UV exactly reproduces the anomaly. The CS coefficient will be fixed to $N_c=3$ when matching QCD. 
The CS $5$-form, defining \mbox{$\A=-i\A_Mdx^M$}, is
\be
\omega_5(\A)\,=\,{\rm Tr}\left[\A (d \A)^2+ \frac{3}{2} \A^3 (d \A) + \frac{3}{5} \A^5\right]\, .
\ee
When $\A$  is the connection of an $U(2)$ group, as in our case,
one can use  the fact that $SU(2)$ is an anomaly-free group to write
$\omega_5$  as
\be
\omega_5(\A)\,=\,\frac32{\widehat{A}}{\rm Tr}\left[F^2\right]+\frac14\widehat{A}\left(d\widehat{A}\right)^2\,+\,
d\,{\rm Tr}\left[\widehat{A} A F -\frac14 \widehat{A} A^3\right] \, ,
\ee
where $\A=A+\widehat{A}\I/2$ and $A$ is the $SU(2)$ connection. 
The total derivative part of the above equation  can be dropped, since it only adds to $S_{CS}$ an UV-boundary term for the sources.

The full 5D action will be given by $S_5=S_g+S_{CS}$, where $S_g$ is made of locally gauge invariant terms. $S_g$ is also invariant under transformations which do not reduce to the identity at the UV-boundary,  and for this reason it does not contribute to the anomalous variation of the partition function. Taking the operators of the lowest dimensionality, 
we have
\beq
S_g=-\int d^4{x}\int^{z_{\IR}}_{z_{\UV}} dz\,  a(z)\, \frac{M_5}{2} \left\{
\Tr\left[{L_{MN}L^{MN}}\right]\,+\,\frac{\alpha^2}2{\widehat{L}}_{MN}{\widehat{L}}^{MN}\,+\,\{L\,\leftrightarrow\,R\}\right\}\, .
\label{Sg}
\eeq
We have imposed  on the 5D theory   invariance under the combined $\{{\bf x}\rightarrow-{\bf x},L\,\leftrightarrow\,R\}$, where ${\bf x}$ denotes ordinary $3$-space coordinates. This symmetry, under which $S_{CS}$ is also invariant, corresponds to the usual parity on the 4D side. 
We 
have normalized differently the kinetic term
of the $SU(2)$ and  $U(1)$ gauge bosons, since 
we do not have any symmetry  reason to put them equal.
In the large-$N_c$ limit of QCD, however, the Zweig's rule leads to equal couplings (and masses) for the $\rho$ and $\omega$ mesons, implying $\alpha=1$ in our 5D model.
Since this well-known feature of large-$N_c$ QCD
 does not arise automatically in our 5D framework (as, for instance, the equality of the $\rho$ and $\omega$ masses does), we will keep $\alpha$ as a free parameter.
The CS term, written in component notation, will be given by
\bea
S_{CS}\,=\,\frac{N_c}{16\pi^2}\int d^5x\left\{
\frac14\epsilon^{MNOPQ}{\widehat{L}_M}\Tr\left[L_{NO}L_{PQ}\right]\right.\nn\\
\,+\,\left.
\frac1{24}\epsilon^{MNOPQ}{\widehat{L}_M}{\widehat{L}_{NO}}{\widehat{L}_{PQ}}\,-\,\{L\,\leftrightarrow\,R\}
\right\}\, .
\label{Scs}
\eea
The 5D theory defined above has only 3 independent parameters: 
$M_5$, $L$ and $\alpha$. 

Let us make again use of Eq.~(\ref{corr}) to determine the current operators 
through which the theory couples to the external EW bosons.
These currents are obtained by varying  Eq.~(\ref{corr}) with respect to  $l_\mu$
 (exactly the same would be true for $r_\mu$) and then taking $l_\mu=r_\mu=0$. 
The variation of the l.h.s. of Eq.~(\ref{corr}) simply gives the current correlator of the 4D theory, 
while in the r.h.s. this corresponds to a   variation of  the UV-boundary conditions. 
The effect of this latter can be calculated in the following way.
We   perform a  field redefinition $\Le_\mu\rightarrow \Le_\mu+\delta \Le_\mu$ where $\delta \Le_\mu(x,z)$  is chosen to  respect the IR-boundary conditions and  
fulfill  $ \delta \Le_\mu(x,z_{\UV})=\delta l_\mu$.  
This redefinition removes the original variation of the UV-boundary conditions, but leads 
a new term  in the  5D action, $\delta S_5$. One then has
\be
i\int d^4x{\textrm{Tr}}\left[\langle j_{L}^\mu(x)\rangle\delta l_\mu(x)\right]\,=\,
i\int \mathcal{D}\Le_M\mathcal{D}\R_M\delta S_5\left[\Le,\R\right]\exp{\left[iS_5\left[\Le,\R\right]\right]}\, ,
\ee
where the 5D path integral is now performed by taking  $l_\mu=r_\mu=0$, {\it{i.e.}} normal Dirichlet 
 conditions. 
 The explicit value  of  $\delta S_5$  is given by
\be
\delta S_5\,=\,\int d^4x {\textrm{Tr}}\left[ \J_{L}^\mu(x)\delta l_\mu(x)\right]\,+\,\int d^5x (\textrm{EOM})\cdot \delta L\, ,
\label{currdef}
\ee
where $\J_{L\, \mu}=J_{L\, \mu}^{a}\sigma^a+\widehat J_{L\, \mu} \I$ and 
\be
J_{L\, \mu}^{a}\,=\,M_5\big(a(z)L_{\mu\,5}^a\big)\left|_{z=z_{\UV}}\right.\ ,
\;\;\;\;\;{\widehat J}_{L\, \mu}\,=\,\alpha^2M_5\big(a(z){\widehat L}_{\mu\,5}\big)\left|_{z=z_{\UV}}\right.\, .
\label{cur0}
\ee
The last term of Eq.~(\ref{currdef}) corresponds to the  5D ``bulk" part of the variation, which leads to the equations of motion (EOM). 
Remembering that the EOM always have zero expectation value \footnote{We have actually shown this here; notice that $\delta \Le_\mu$ was completely arbitrary in the bulk, but the variation of the functional integral can only depend on $\delta l_\mu = \delta \Le_\mu (x,z_{\UV})$.}, 
we find that we can identify $\J_{L}^\mu$ of  Eq.~(\ref{cur0}) with the current operator on the 5D side: $\langle j_L^\mu\rangle_{\rm 4D}=\langle \J_L^\mu\rangle_{\rm 5D}$.
Notice that the CS term has  not contributed to Eq.~(\ref{currdef}) due to the fact that each term in $S_{CS}$ which contains a $\partial_z$ derivative (and therefore could lead to a UV-boundary term) also contains  $\Le_\mu$ or $\R_\mu$  fields;
these fields  on   the UV-boundary  are the sources $l_\mu$ and $r_\mu$ 
that must be put to zero.

\subsection{Meson Physics and Calculability}

The phenomenological implications 
for the lightest mesons  
 of  5D  models like the one described above 
have been extensively studied  in the literature.
Let us briefly  summarize the main results here. If rewritten in 4D terms, the theory contains massless  Goldstone bosons that
parametrize the $U(2)_L\times U(2)_R/U(2)_V$ coset and describe the pion triplet and a massless  $\eta'$. The pion decay constant is given by
\begin{equation}
\label{fpi}
F_\pi^2=2 M_5\left(\int \frac{dz}{a(z)}\right)^{-1}=\frac{4 M_5}{L}\, .
\end{equation}
 The massive spectrum consists of infinite 
towers of vector and axial-vector spin-one KK resonances. Among the vectors we have an 
isospin triplet, the $\rho^{(n)}$, and a singlet $\omega^{(n)}$. The axial-vectors are
again a triplet $a_1^{(n)}$ and a singlet $f_1^{(n)}$. We want to interpret, as our 
terminology already suggests, the lightest states of each tower as the $\rho(770)$, 
$\omega(782)$, $a_1(1260)$ and $f_1(1285)$ resonances, respectively.
The model predicts  at leading order, {\it i.e.} at tree-level,
\begin{equation}
\label{masses}
m_\rho=m_\omega\simeq \frac{3\pi }{4L}\ , \ \ \  m_{a_1}=m_{f_1}\simeq \frac{5\pi}{4L}\, ,
\end{equation}
compatibly  with  observations. 
%vector and the two axial--vector towers to be degenerate in mass. 
%For an AdS$_5$ warp factor as in Eq.~(\ref{warpf}) the vector and axial-vector masses 
%are provided by certain non-analitic functions $m_V(L)$ and $m_A(L)$ whose exact form 
%is irrelevant here, we just want to point out that the masses are independent of the 
%5D coupling $M_5$ and of the parameter $\alpha$. What crucially depends on $M_5$ are 
%of course the mesons couplings like $g_{\rho\pi\pi}$, which scales like $1/\sqrt{M_5}$, 
%and the decay constants $f_M\propto \sqrt{M_5}$ where $M=\rho,a_1,\omega$. 
%\footnote{We are thinking, differently from our conventions in \cite{}, to the standard definition
%of the decay constants as
%$$
%\langle 0 |J_\mu |M\rangle = \epsilon_\mu m_M f_M\,,
%$$
%where $J$ represents the appropriate current and $\epsilon_\mu$ and $m_M$ are the polarization vector and the mass of the meson.} 
%The parameters $\alpha$ controls the $\omega$ and $\rho$ decay constant ratio, 
%$f_\omega/f_\rho=\alpha$, and the CS--induced decay rates of the $\omega$ discussed in \cite{}.
The model also predicts the decay constants $F_i$ and couplings $g_i$ for the mesons as a function of $M_5$, $L$  and  $\alpha$ that can be found in  Refs.~\cite{Erlich:2005qh,Rold:2005,Hirn:2005nr,Pomarol:2007kr}; here 
we only  notice, for later use,  their  scaling with the 5D coupling: 
\begin{equation}
\label{couplings}
F_i\sim \sqrt{M_5}\ ,\ \ \ \  g_i  \sim \frac{1}{\sqrt{M_5}}\, ,
\end{equation}
while the masses, as shown above, do not  depend on $M_5$.
In Table~\ref{pretotal} we show
a fit to $14$ meson quantities.  The best fit is obtained for the values of $1/L=343$ MeV, $M_5 L=0.0165$ and  $\alpha=0.94$ for the three 
parameters of our model. The minimum Root Mean Square Error (RMSE) corresponding to those values 
is found to be $11\%$ and the relative deviation of each single prediction is below around 
$15\%$.

\begin{table}[tp]
%\small
\tbl{Global fit to mesonic   physical quantities. 
Masses, decay constants and widths  are given in MeV.
Physical masses have been used in the kinematic factors of the partial decay  widths.}
{\begin{tabular}{ccccc}
\hline
  & {\rm Experiment}
    &{\rm AdS$_5$}& Deviation
\\ \hline
 $m_\rho$ & $775 $ & $824$ & $+6\%$
\\
 $m_{a_1}$ & $1230 $ &  $1347$    &+$10\%$
\\
 $m_\omega$ & $782 $ & $824$ &$+5\%$
\\
 $F_\rho$ & $153$ & $169$ &$+11\%$
\\
 $F_{\omega}/F_\rho$ & $0.88$   &$0.94$&$+7\%$
\\ 
 $F_{\pi}$ & $87$ &    $88$& $+1\%$
\\ 
 $g_{\rho\pi\pi}$ & $6.0$ & $5.4$& $-10\%$
\\
 $L_9$ & $6.9\cdot 10^{-3}$ &     $6.2\cdot 10^{-3}$&$-10\%$
\\
 $L_{10}$ & $-5.2\cdot 10^{-3}$ &    $-6.2\cdot 10^{-3}$&$-12\%$
\\
 $\Gamma(\omega\rightarrow \pi\gamma)$ & $0.75$   & $0.81$
           &$+8\%$
            \\
 $\Gamma(\omega\rightarrow 3\pi)$ & $7.5$   & $6.7$      &$-11\%$
            \\
 $\Gamma(\rho\rightarrow \pi\gamma)$ & $0.068$   & $0.077$          &$+13\%$
 \\
 $\Gamma(\omega\rightarrow \pi\mu\mu)$ & $8.2\cdot 10^{-4}$   & $7.3\cdot 10^{-4}$      &$-10\%$
 \\
 $\Gamma(\omega\rightarrow \pi e e )$ & $6.5\cdot 10^{-3}$   & $7.3\cdot 10^{-3}$         &$+12\%$
           \\
\hline
\end{tabular}}
\label{pretotal}
\end{table}

Concerning the choice of the meson observables, some remarks are in order. First of all, 
we are only considering the lowest state of  each KK tower  because we expect the masses and 
couplings of the heavier mesons to receive large quantum corrections. Our model is indeed, 
as we will explain below, an effective theory valid up to a cut-off 
$\Lambda_5\sim 2$ GeV  and our tree-level calculations only correspond to the leading term of an $E/\Lambda_5$ expansion. 
%A  comparison could be at most attempted for the second KK of the $\rho$, the $\rho(1450)$ resonance.
%We would obtain a fairly good prediction of its mass, a calculation of its decays has not 
%been performed yet. 
%Our theory also contains additional scalar resonances, which correspond to the 
%longitudinal (in a 5D sense) components of the 5D gauge fields. They are completely decoupled, of 
%course, as long as the gauge symmetry is exact in the 5D bulk, but they become relevant as soon as the 5D symmetry 
%is broken to account for the explicit breaking of the chiral symmetry. Those resonances 
%have been studied in Ref.~\cite{Rold:2005}, we ignore them because we are working, for simplicity, with exact  chiral symmetry. 
Apart from this restriction, we must   include  in our fit  observables
 with an  experimental  accuracy better than $10\%$. 
This is because we want  to neglect the experimental error in order to obtain an 
estimate of the accuracy of our theoretical predictions. 
Much more observables can be computed, once the best-fit value of the parameters are 
obtained, and several of them have already been considered in the literature.
For instance, one can study the other low-energy constants of the chiral lagrangian, 
 the physics of the  $f_1$ resonance or  the pseudo--scalar resonances
which arise when the explicit breaking of the chiral symmetry is taken into account
\cite{Rold:2005}. It would also be interesting to compute 
the $a_1\rightarrow \pi\gamma$  decay, which is absent in our model at tree-level and 
only proceeds via loop effects or higher-dimensional terms of our 5D effective lagrangian. 
\footnote{Higher order contributions will also change our tree-level prediction 
$L_9+L_{10}=0$, which is again related with the 
absence of the $a_1$--$\pi$--$\gamma$ vertex.}

As discussed in the Introduction, the semiclassical expansion in the 5D 
model should correspond to the large-$N_c$ expansion on the 4D side.
The results presented above provide a confirmation of this 
interpretation: at large-$N_c$ meson masses are expected to scale like $N_c^0$, while meson couplings  and decay constants scale like $g_{i},1/F_i\sim 1/\sqrt{N_c}$. 
These scalings agree with Eq.~(\ref{masses}) and (\ref{couplings})
if the parameters $\alpha$, $L$  and $M_5$ are taken to scale like 
\footnote{This scaling can also be obtained from the AdS/CFT correspondence.}
\begin{equation}
\label{scaling}
\alpha\sim N_c^0\ ,\ \ L\sim N_c^0\ ,\ \  M_5\sim N_c\, .
\end{equation}
This leads us to define the adimensional $N_c$-invariant parameter
\be
\label{gamma}
\gamma\equiv \frac{N_c}{16\pi^2 M_5 L\alpha}\, ,
\ee
whose experimental value is $\gamma=1.23$ and will be useful later on.
We will also show in the following that the assumed scaling 
of the 5D  parameters leads to the correct $N_c$ scaling  in the baryon sector as well.

Other descriptions of vector mesons in terms of massive vector fields, {\it i.e.} models with 
 Hidden Local Symmetry (HLS) \cite{Bando:1987br,Georgi:1989xy} or two-form fields
 \cite{Ecker:1989yg}, also
 correctly reproduce the  meson physical properties.
Nevertheless,  we  believe that 5D  models,  as the one discussed here, 
present more advantages.
\footnote{It must be possible, generalizing what was done in Ref.~\cite{Son:2003et}, to rewrite 
our model as a 4D HLS with infinitely many $U(2)$ hidden  symmetry groups. The comparison with
HLS models that we  perform in this section only applies, therefore, to the standard case 
of a finite number of hidden symmetries.} First of all, they contain less parameters.  
In  the  models of Refs.~\cite{Bando:1987br,Georgi:1989xy,Ecker:1989yg}, for example,  the mass and the couplings of  each meson are independent parameters;
 also anomalous processes, those involving an odd 
number of pions, 
depend  on several operators with unknown coefficients which arise at the same order,
  while in our case  they all arise from a single operator, the 5D CS term. 
Finally, Vector Meson Dominance is automatic in our scenario, while it needs to be imposed ``by hand'' in the case of HLS.

Moreover, and perhaps more importantly, 5D models  are  calculable effective field theory 
in which higher-dimensional operators are suppressed by the 
cut-off of the theory $\Lambda_5$. Calculations can be organized as an expansion in 
$E/\Lambda_5$, where $E$ is the typical scale of the process under consideration.
Given that the cut-off is parametrically 
bigger than the mass of the lightest mesons, reliable calculations 
of masses and couplings can be performed. 
%The situation is different in HLS models, where the cut-off  is $\Lambda_4\sim 4\pi f_\pi\sim m_\rho$.

Let us now use naive dimensional arguments to estimate the maximal value of our 
cut-off $\Lambda_5$. This is determined by the scale
at which   loops   are of order  of tree-level effects.
Computing loop corrections to the $F^2$ operator, which arise from 
the $F^2$ term itself, one gets  $\Lambda_5\sim 24\pi^3 M_5$. 
Nevertheless, one gets a lower value for $\Lambda_5$ from the CS term
due to the $N_c$ dependence of its coefficient.
Indeed, at the one-loop level, the CS term gives a contribution of order $M_5$
to the $F^2$ operator for
$\Lambda_5\sim 24\pi^3M_5/N_c^{2/3}$.
Even though  the cut-off scale  lowers due to the presence of the CS term,  we can still have, in the large-$N_c$ limit,
a 5D weakly coupled theory where higher-dimensional operators are suppressed. 
The cut-off can be rewritten as
$$
\Lambda_5\sim\frac{3\pi}2\frac{{N_c}^{1/3}}{\gamma\alpha L}\sim 2\ \text{GeV}\,,
$$
where we have used the best-fit value of our parameters. 

The power of calculability  of our  5D model
makes it   very suitable for  studying  baryon physics.
Indeed, the typical size of the 5D skyrmion solution will  be  of order $\rho_s\sim 1/m_\rho$,
guaranteeing that effects  from higher-dimensional operators  
will be suppressed by  $m_\rho/\Lambda_5\sim 0.4$.
This  is therefore, we believe,  the first fully consistent approach   towards  baryon physics.

\section{Baryons from 5D Skyrmions}
\label{bar}
%Define Baryons as solitons in 5D. Topology.
%Relation with Skyrmions. Relation with the large-N limit.
%Equations and bc.

%AA2+AG(Andrea-Giuliano) primera parte

\subsection{4D Skyrmions from 5D Solitons}

Time-independent configurations of our 5D fields, which correspond to 
allowed initial ($t\rightarrow-\infty$) and final ($t\rightarrow+\infty$) states 
of the time evolution, are labeled by the topological charge
\beq
B=\frac1{32\pi^2}\int d^3x\int^{z_{\IR}}_{z_{\UV}} dz\,  
\epsilon_{\hat\mu\hat\nu\hat\rho\hat\sigma}\Tr\left[{L^{\hat\mu\hat\nu}L^{\hat\rho\hat\sigma}}
-{R^{\hat\mu\hat\nu}R^{\hat\rho\hat\sigma}}\right]\,,
\label{Bch}
\eeq
where the indeces $\hat\mu,\hat\nu,\ldots$ run over the $4$ spatial coordinates, 
but they are raised with Euclidean metric.
We will now show that $B$ can only assume integer values, which ensures that 
it cannot be changed by the time evolution. This makes $B$ a topologically conserved 
charge which we identify with the baryon number. 
In order to show this, and with the aim of making the relation with the skyrmion more precise, it is convenient to go to the axial gauge 
$L_5=R_5=0$. The latter can be easily reached, starting from a generic gauge field configuration, by means of a Wilson-line transformation. 
In the axial gauge both boundary conditions Eqs.~(\ref{irboundary condition}) and (\ref{uvboundary condition}) (in which we take now $l=r=0$) cannot be
simultaneously satisfied. Let us  then 
keep  Eq.~(\ref{irboundary condition})  but modify  the UV-boundary condition to
\beq
{\widetilde L}_{i}\left|_{z=z_{\UV}}\right.=i\,U({\mathbf x})\partial_iU({\mathbf x})^\dagger\,,\;\;\;\;\;{\widetilde R}_{i}\left|_{z=z_{\UV}}\right.=0\,,
\label{uvnboundary condition}
\eeq
where $\widetilde L_i$ and $\widetilde R_i$ are the gauge fields in the axial gauge and 
$i$ runs over the $3$ ordinary space coordinates.  The field $U({\mathbf x})$ in the equation above precisely corresponds to the Goldstone field in the 4D interpretation of the model 
\cite{Panico:2007qd}.
%By using a form notation \cite{Chu:1996fr} $A=-i\,A_{\hat\mu} dx^{\hat\mu}$ and 
 Remembering that $F\wedge F=d\omega_3$, 
where $\omega_3$ is the third   CS form, 
 the 4D integral in Eq.~(\ref{Bch}) can be rewritten as an integral 
%of the third Chern-Simons form $\omega_3(A)$
 on the $3D$ boundary of the space:
 \beq
B\,=\,\frac1{8\pi^2}\int_{3D}\left[\omega_3({\widetilde L}) -
\omega_3({\widetilde R})\right]\, .
\label{qw3}
 \eeq
   The contribution to $B$ coming from  the IR-boundary
 vanishes as the $L$ and $R$ terms in Eq.~(\ref{qw3}) cancel each other due to Eq.~(\ref{irboundary condition}).
  This is crucial for $B$ to be quantized and it is the reason why we have to choose the relative minus sign among the $L$ and $R$ instanton charges in the definition of $B$. 
At the ${\mathbf x}^2\rightarrow\infty$ boundary, the contribution to $B$ also vanishes
since in the axial gauge $\partial_5 A_i=0$ (in order to have $F_{5i}=0$).    
We are then  left with the UV-boundary  which we can topologically regard
as the $3$-sphere $S_3$. Therefore, we find
\bea
B\,=&&\,-\frac1{8\pi^2}\int_{\UV}\omega_3\left[{\widetilde L}_i\left(=i\,U\partial_iU^\dagger\right)
\right]\nn\\
\,=&&\,\frac1{24\pi^2}\int d^3x\, \epsilon^{ijk}{\textrm Tr}\left[
U\partial_iU^\dagger\,U\partial_jU^\dagger\,U\partial_kU^\dagger\right]\,\,\in\ZZ\,.
\label{qfinal}
\eea
The charge $B$ is equal to the Cartan-Maurer integral invariant for $SU(2)$ which is an integer.
% and corresponds to the Baryon number $B$ in the Skyrme model. 

In the next section we will discuss regular static  solutions with nonzero $B$.
If they exist, they cannot trivially correspond to a pure gauge configuration.
% and they must have positive energy. 
Moreover, the particles associated to solitons with $B=\pm1$ will be
 stable given that they have minimal charge.
 Eq.~(\ref{qfinal}) also makes the relation with 4D skyrmions explicit: 
 topologically non-trivial 5D configurations are 
  those for which the corresponding pion matrix
  $U({\mathbf x})$ is also non-trivial. The latter 
 corresponds to a 4D skyrmion
 with  baryon number $B$. 
In  a general gauge, the skyrimion configuration  $U({\mathbf x})$   will be given by
\beq
U({\mathbf x})\,=\,P\left\{\exp\left[-i\,\int_{z_{\UV}}^{z_{\IR}}dz'\,R_5({\mathbf x},z')\right]\right\}\cdot P\left\{\exp\left[i\,\int_{z_{\UV}}^{z_{\IR}}dz'\,L_5({\mathbf x},z')\right]\right\}\, ,
\label{sky}
\eeq
where $P$ indicates   path ordering.  
From  a 4D  perspective, the 5D soliton that we are looking for 
can be considered to be a 4D skyrmion made
of Goldstone bosons and  the  massive tower of KK gauge bosons.

\subsection{The Static Solution}
\label{ss}

In order to obtain the static soliton solution of the 5D EOM of our theory it is crucial to 
specify a correct Ansatz, which is best constructed by exploiting the symmetries of our 
problem. Let us impose, first of all, our solution to be invariant under 
 time-reversal $t\rightarrow-t$ combined with ${\widehat{L}}\rightarrow-{\widehat{L}}$ and 
 ${\widehat{R}}\rightarrow-{\widehat{R}}$, under which also  the CS term is invariant.  
 This transformation reduces,
in static configurations,  to a sign change of the temporal component of $L$ and $R$ and of the 
spatial components of ${\widehat{L}}$ and ${\widehat{R}}$. 
We can therefore consistently put them to zero. 
We also use parity invariance 
(\mbox{$\{ L\leftrightarrow R, {\bf  x}\leftrightarrow -{\bf x}\}$})
 to restrict to configurations for which
$L_i({\bf x},z,t)=-R_i(-{\bf x},z,t)$, $L_{5,0}({\bf  x},z,t)=R_{5,0}(-{\bf x},z,t)$ and analogously
for ${\hat L}$, ${\hat  R}$.
We impose, finally, invariance under ''cylindrical'' transformations \cite{Witten:1976ck}, 
{\it i.e.}  the simultaneous action of 3D space rotations 
$x_a\sigma^a\,\rightarrow\,\theta^\dagger x_a\sigma^a\theta$,
with $\theta\in SU(2)$, and vector $SU(2)$ global transformations $L,R\rightarrow \theta\,(L,R)\,\theta^\dagger$. 
An equivalent way to
state the invariance is that a 3D rotation with $\theta$ acts on 
the solution exactly as an $SU(2)$ 
vector one in the
opposite direction ({\it i.e.} with $\theta^\dagger$) would do.
The resulting Ansatz for the static solution (which we denote by ``barred'' fields)
 is entirely specified  $4$ real 2D fields
\be
\left\{
\begin{array}{l}
\displaystyle
{\ov R}^a_j({\bf x},z) = \displaystyle A_1(r,z) \x_a \x_j + \frac1r \varepsilon_{ajk} \x_k
-\frac{\phi_{(x)}}r\varepsilon^{(x,y)}\Delta^{(y),aj} \,,\\
\displaystyle
{\ov R}^a_5({\bf x},z) = \displaystyle  A_2(r,z)\x^a \,,\\
\displaystyle
\alpha\widehat {\ov R}_0({\bf x},z) = \displaystyle \frac{s(r,z)}r \,,
\end{array}
\right.
\label{sts}
\ee
where $r^2=\sum_i x^i x^i$, \  $\x^i=x^i/r$, $\varepsilon^{(x,y)}$ is the antisymmetric tensor with
$\varepsilon^{(1,2)}=1$ and the ``doublet'' tensors $\Delta^{(1,2)}$ are
\be
\Delta^{(x),ab}\,=\, \left[
\begin{array}{l}
 \epsilon^{abc}\x^c\\
\x^a\x^b - \delta^{ab}
\end{array}
\right]\,.
\label{defde}
\ee

Substituting the Ansatz in the topological charge Eq.~(\ref{Bch}) we find
\beq
B=\frac{1}{2\pi}\int_{0}^\infty dr\int^{z_{\IR}}_{z_{\UV}}dz\,\epsilon^{\bar\mu\bar\nu}
\bigg[\partial_{\bar\mu}(-i \phi^*D_{\bar\nu}\phi+h.c.)
+A_{\bar\mu\bar\nu}\bigg]\,,
\label{topcharge2d}
\eeq
where $x^{\bar\mu}=\{r,z\}$, 
$A_{\bar\mu}=\{A_1,A_2\}$, $A_{\bar\mu\bar\nu}$ its field-strenght, 
$\phi=\phi_1+i\phi_2$ and the covariant derivative will be defined in Eq.~(\ref{cder}).
The charge can be written, as it should, as an integral over the 1D boundary of the 2D space. 
Finite-energy regular solutions with $B=1$ which obey Eqs.~(\ref{irboundary condition}) 
and (\ref{uvboundary condition}) must respect the following boundary conditions:
\be
z=z_{\IR}\ :
\quad
\left\{
\begin{array}{l}
\phi_1 = 0\\
\partial_2 \phi_2 = 0\\
A_1 = 0\\
\partial_2 s = 0
\end{array}
\right.\,,
\qquad\qquad
z=z_{\UV}\ :
\quad
\left\{
\begin{array}{l}
\phi_1 = 0\\
\phi_2 = -1\\
A_1 = 0\\
s=0
\end{array}
\right.\,,
\qquad\qquad
\label{bco}
\ee
and
\be
r=0\ :
\quad
\left\{
\begin{array}{l}
\phi_1/r \rightarrow A_1\\
(1+\phi_2)/r \rightarrow 0\\
A_2 = 0\\
s=0
\end{array}
\right.
\quad\qquad
r=\infty\ :
\quad
\left\{
\begin{array}{l}
\phi = -i e^{i \pi z/L}\\
A_2 = \frac{\pi}{L}\\
s=0
\end{array}
\right.\,.
\label{bco1}
\ee
Solutions of the EOM with the required boundary conditions exist, and have been obtained 
numerically in Ref.~\cite{Pomarol:2008aa} using  the COMSOL package \cite{comsol}
(see Appendix for details).
The 2D energy density of this solution is given in Fig.~\ref{bola}.

\subsection{Zero-Mode Fluctuations}

Let us now consider time-dependent infinitesimal deformations of the static solutions. Among these, the zero-mode
({\it i.e.} zero frequency) fluctuations are particularly important as they will describe single-baryon states.
Zero-modes can be defined as directions in the field space in which uniform and slow motion is permitted by the
classical dynamics and they are associated with the global symmetries of the problem, which are in our case $U(2)_V$
and $3$-space rotations plus $3$-space translations. The latter would describe baryons moving with uniform velocity
and therefore can be ignored in the computation of static properties like the form factors. Of course, the global
$U(1)_V$ acts trivially on all our fields and the global $SU(2)_V$ has the same effect as $3$-space rotations on
the static solution (\ref{sts}) because of the cylindrical symmetry. The space of static solutions which are
of interest for us is therefore parametrized by $3$ real coordinates --denoted as collective coordinates--
which define an $SU(2)$ matrix $U$.

To construct zero-modes fluctuations we consider collective coordinates with general time dependence, {\it i.e.}
we perform a global $SU(2)_V$ transformation on the static solution
\be
R_{\hat\mu}({\bf x},z;U)\,=\,U\,{\ov R}_{\hat\mu}({\bf x},z)\,U^\dagger\  ,
\;\;\;\;\;
{\widehat{R}_0}({\bf x},z;U)\,=\,{{{\widehat{\ov{R}}}}}_0({\bf x},z)\,,
\label{eq0m}
\ee
but we allow $U=U(t)$ to depend on time. It is only for constant $U$ that Eq.~(\ref{eq0m}) is a solution of
the time-dependent EOM. For infinitesimal but non-zero rotational velocity
$$
K=k_a\sigma^a/2=-i U^\dagger dU/dt\,,
$$
Eq.~(\ref{eq0m}) becomes an infinitesimal deformation of the static solution. Along the zero-mode direction
uniform and slow motion is classically allowed, for this reason our fluctuations should fulfill the
time-dependent EOM at linear order in $K$ provided that $d K/dt=0$.

From the action (\ref{Sg}) and (\ref{Scs}) the following EOM are derived
\be
\left\{
\begin{array}{l}
\displaystyle
D_{\hat\nu}\left(a(z)R^{\hat\nu}_{\;0}\right)+\frac{\gamma  \alpha L}4\epsilon^{\hat\nu\hat\omega\hat\rho\hat\sigma}R_{\hat\nu\hat\omega}{\widehat R}_{\hat\rho\hat\sigma}=0  \\
\displaystyle
\alpha\partial_{\hat\nu}\left(a(z){\hat R}^{\hat\nu}_{\;\  0}\right)+\frac{\gamma  L}4\epsilon^{\hat\nu\hat\omega\hat\rho\hat\sigma}\left[{\rm  Tr}\left(R_{\hat\nu\hat\omega}R_{\hat\rho\hat\sigma}\right)  +\frac12{\widehat R}_{\hat\nu\hat\omega}{\widehat  R}_{\hat\rho\hat\sigma}\right]=0\\
\displaystyle
D_{\hat\nu}\left(a(z)R^{\hat\nu\hat\mu}\right)-a(z)D_0R_{0}^{\;\  \hat\mu}-\frac{\gamma\alpha   L}2\epsilon^{\hat\mu\hat\nu\hat\rho\hat\sigma}\left[R_{\hat\nu  0}{\widehat R}_{\hat\rho\hat\sigma}+R_{\hat\nu \hat\rho}{\widehat  R}_{\hat\sigma 0}\right]=0\\
\displaystyle
\alpha\partial_{\hat\nu}\left(a(z){\widehat  R}^{\hat\nu\hat\mu}\right)-\alpha a(z)\partial_0{\widehat R}_{0}^{\;\  \hat\mu}-\gamma L\epsilon^{\hat\mu\hat\nu\hat\rho\hat\sigma}\left[{\rm  Tr}\left(R_{\hat\nu 0} R_{\hat\rho\hat\sigma}\right)+\frac12{\widehat  R}_{\hat\nu 0}{\widehat R}_{\hat\rho\hat\sigma}\right]=0
\end{array}
\right. \, .
 \label{eomt}
 \ee
We only need to specify the EOM for one chirality since we are considering, as explained in the previous section,
a parity invariant Ansatz. We would like to find solutions of Eq.~(\ref{eomt}) for which  $R_{\hat\mu}$ and
${\widehat R}_{0}$ are of the form (\ref{eq0m}); it is easy to see that the  time-dependence of $U$ in
Eq.~(\ref{eq0m}) acts as a source for the components $R_{0}$ and ${\widehat  R}_{\hat\mu}$, which therefore cannot
be put to zero as in the static case. Notice that the same happens in the case of the 4D  skyrmion
\cite{Meissner:1987ge}, in which  the temporal and spatial components of the $\rho$ and $\omega$ mesons are turned
on in the rotating skyrmion solution. Also, it can be shown that Eq.~(\ref{eomt}) can be solved, to linear order
in $K$ and for $dK/dt=0$, by the Ansatz in Eq.~(\ref{eq0m}) if the fields $R_{0}$ and ${\widehat R}_{\hat\mu}$ are
chosen to be linear in $K$.
Even though $K$ must be constant for the EOM to be solved, it should be clear that this
does not imply any constraint on the allowed form of the collective coordinate matrix
$U(t)$ in Eq.~(\ref{eq0m}), which can have an arbitrary dependence on time. What we actually
want to do here is to find an appropriate functional dependence of the fields on $U(t)$
such that the time-dependent EOM would be solved if and only if the rotational velocity
$K=-i U^\dagger dU/dt$ was constant.

In order to solve the time-dependent equations (\ref{eomt}) we will consider a 2D Ansatz obtained by a
generalization of the cylindrical symmetry of the static case. The Ansatz for $R_{\hat\mu}$ and
${\widehat R}_{0}$ is specified by Eq.~(\ref{eq0m}) in which the static fields are given by Eq.~(\ref{sts}).
Due to the cylindrical symmetry of the static solution the fields in Eq.~(\ref{eq0m}) are invariant under 3D space
rotations $x_a\sigma^a\,\rightarrow\,\theta^\dagger x_a\sigma^a\theta$ combined with vector $SU(2)$ global transformations
$L,R\rightarrow \theta\,(L,R)\,\theta^\dagger$ if $U$ also transforms as 
$U\rightarrow \theta^\dagger U \theta$. We are therefore led
to consider a generalized cylindrical symmetry under which $k_a$ also rotates as the space coordinates do.
Compatibly with this symmetry and with the fact that $R_{0}$ and ${\widehat R}_{\hat\mu}$ must be linear in $K$
we write the Ansatz as
\be
\displaystyle
R_{0}({\bf x},z;U)\,=\,U\,{\ov R}_{0}({\bf x},z;K)\,U^\dagger\,+\,i\,U\partial_0U^\dagger \, ,
\;\;\;\;\;
{\widehat{R}_{\hat\mu}}({\bf x},z;U)\,=\,{{{\widehat{\ov{R}}}}}_{\hat\mu}({\bf x},z;K)\,,
\label{ansk}
\ee
where
\be
\left\{
\begin{array}{l}
{\ov R}_{0}^a({\bf x},z;K) =\displaystyle \chi_{(x)}(r,z)k_b\Delta^{(x),ab} + v(r,z) (k\cdot\x)\x^a\,\\
\displaystyle
\alpha{{{\widehat{\ov{R}}}}}_{i}({\bf x},z;K) = \displaystyle \frac{\rho(r,z)}{r}\left(k^i - (k\cdot\x)\x^i\right) + B_1(r,z)(k\cdot\x)\x^i + Q(r,z)\epsilon^{ibc}k_b\x_c \,\\
\displaystyle
\rule{0pt}{1.5em}\alpha{{{\widehat{\ov{R}}}}}_{5}({\bf x},z;K) = \displaystyle B_2(r,z)(k\cdot\x) \,
\end{array}
\right.\,.
\label{ansk1}
\ee

It must be observed that our Ansatz has not fixed the 5D gauge freedom completely; its 
form is indeed preserved
by chiral $SU(2)_{L,R}$ gauge transformations of the form
$g_R=U(t)\cdot g\cdot U^\dagger (t)$ and $g_L=U(t)\cdot g^\dagger\cdot U^\dagger (t)$ with
\be
g = \exp[i \alpha(r,z) x^a \sigma_a/(2r)]\,,
\label{eq:resU1SU2}
\ee
under which the 2D fields $\phi_{(x)}$ and $\chi_{(x)}$ defined respectively in Eq.~(\ref{sts}) and (\ref{ansk1})
transform as charged complex scalars.
The fields
$A_{\bar\mu}$ transform as gauge fields.
There is also a second residual $U(1)$ associated with chiral $U(1)_{L,R}$ 5D transformations of the form
${\widehat g}_R={\widehat g}$ and ${\widehat g}_L={\widehat g}^\dagger$ with
\be
{\widehat g} = \exp\left[i \beta(r,z)\frac{(k\cdot\x)}\alpha\right]\,.
\label{eq:resU1U1}
\ee
Under this second residual $U(1)$ only $B_{\bar \mu}=\{B_1,B_2\}$ and $\rho$ transform non 
trivially;
$B_{\bar \mu}$ is a gauge field and $\rho$ a Goldstone.
In order to make manifest the residual gauge invariance of the observables
we will compute we introduce gauge covariant derivatives for the $\phi$, $\chi$
and $\rho$ fields
\be
\left\{
\begin{array}{l}
\displaystyle
(D_{\bar \mu} \phi)_{(x)} = \partial_{\bar \mu} \phi_{(x)}
+ \epsilon^{(xy)} A_{\bar\mu} \phi_{(y)}\\
\displaystyle
(D_{\bar \mu} \chi)_{(x)} = \partial_{\bar \mu} \chi_{(x)}
+ \epsilon^{(xy)} A_{\bar\mu} \chi_{(y)}\\
\displaystyle
D_{\bar \mu} \rho = \partial_{\bar \mu } \rho - B_{\bar \mu}
\end{array}
\right.\,.
\label{cder}
\ee

At this point it is straightforward to find the zero-mode solution. The EOM for the 2D fields can be obtained
by plugging the Ansatz in Eq.~(\ref{eomt}), while the conditions at the IR and UV boundaries are derived
from Eq.~(\ref{irboundary condition}) and (\ref{uvboundary condition}), respectively. The boundary conditions at $r=0$
are obtained by imposing the regularity of the Ansatz, while those for $r\rightarrow\infty$
come from requiring the energy
of the solution to be finite and $B=1$.
Also in this case, numerical solutions can be obtained with the methods discussed in the appendix. 
The reader not interested in detail can simply accept that a solution
of Eq.~(\ref{eomt}) exists and is given by our Ansatz for some particular
functional form of the 2D fields which we are able to determine numerically. In
the rest of the paper the 2D fields will always denote this numerical solution of the 2D
equations.

\subsection{The Lagrangian of Collective Coordinates}

The collective coordinate matrix $U(t)$ will be associated with static baryons.
The classical dynamics of the collective coordinates is obtained by plugging Eqs.~(\ref{eq0m}) and (\ref{ansk}) in the 5D action. One finds $S[U]=\int dt L$ where
\be
L = -M +\frac{\lambda}2\,k_ak^a\,.
\label{eq:Lag}
\ee
The mass $M$ and the moment of inertia $\lambda$ are given respectively by
\be
\begin{array}{l}
\displaystyle
M=8\pi M_5\int_0^\infty dr\int^{z_{\rm IR}}_{z_{\rm UV}}dz\,\left\{a(z)\left[
|D_{\bar\mu}\phi|^2+\frac{1}{4}r^2 A^{2}_{\bar\mu\bar\nu}
+\frac{1}{2r^2}\left(1-|\phi|^2\right)^2-\frac12\left(\partial_{\bar\mu} s\right)^2\right]\right.\\
\displaystyle\left.
-\frac{\gamma L}2
\frac{s}{r}\epsilon^{\bar\mu\bar\nu}
\bigg[\partial_{\bar\mu}(-i \phi^*D_{\bar\nu}\phi+h.c.)
+A_{\bar\mu\bar\nu}\bigg]
\right\}
\,,
\end{array}
\label{eq:mass}
\ee
and
\be
\begin{array}{l}
\displaystyle
\lambda\,=\,16\pi M_5\frac13\int_0^\infty dr\int^{z_{\rm IR}}_{z_{\rm UV}}dz\,\left\{a(z)\left[
-\left(D_{\bar\mu}\rho\right)^2
-r^2\left(\partial_{\bar\mu} Q\right)^2
-2Q^2
-\frac{r^2}4B_{{\bar\mu}{\bar\nu}}B_{{\bar\mu}{\bar\nu}}\right.\right.\\
\displaystyle
\left.\left.
+r^2\left(D_{\bar\mu}\chi\right)^2+
\frac{r^2}2\left(\partial_{\bar\mu}v\right)^2
+\left(\chi_{(x)}\chi_{(x)}+v^2\right)\left(1+\phi_{(x)}\phi_{(x)}\right)
-4v\phi_{(x)}\chi_{(x)}
\right]\right.\\
\displaystyle
\left.
+\gamma L
\bigg[
-2\epsilon^{\bar\mu\bar\nu}D_{\bar\mu}\rho\,\chi_{(x)}\left(D_{\bar\nu}\phi\right)_{(x)}
+2\epsilon^{\bar\mu\bar\nu}\partial_{\bar\mu}\left(r\,Q\right)\,\chi_{(x)}\epsilon^{(xy)}\left(D_{\bar\nu}\phi\right)_{(y)}\right.\\
\displaystyle
\left.
-v\left(\frac12\epsilon^{\bar\mu\bar\nu}B_{\bar\mu\bar\nu}\left(\phi_{(x)}\phi_{(x)}-1\right)
+r\,Q\epsilon^{\bar\mu\bar\nu}A_{\bar\mu\bar\nu}
\right)
+\frac{2r\,Q}{\alpha^2}\epsilon^{\bar\mu\bar\nu} D_{\bar\mu}\rho\partial_{\bar\nu}\left(\frac{s}{r}\right)
\bigg]
\right\}\,.
\end{array}
\label{eq:lambda}
\ee
The numerical values of $M$ and $\lambda$ are easily computed, 
once the numerical solution for the 2D fields is known. Using the best-fit values of the 
parameters we find $M=1132$ MeV and $1/\lambda=227$ MeV.

Let us give some more detail on this theory. For now we proceed at the classical level and we will discuss the quantization in the next section. Our lagrangian can be rewritten as
\be
L\,=\, -M + \lambda {\rm Tr}\left[{\dot{U}}^\dagger {\dot{U}}\right]\,=\,-M\,+\,2\lambda \sum_{i}{\dot u}_{i}^2\,,
\label{ccL}
\ee
where we have parametrized the collective coordinates matrix $U$ as $U=u_0\I+i\,u_i\sigma^i$, with
$\sum_{i}{u_i}^2=1$. The lagrangian (\ref{ccL}) is the one of the classical spherical rigid rotor. The variables
$\{u_0,u_i\}$ are restricted to the unitary sphere $S^3$, which is conveniently parametrized by the coordinates
$q^\alpha\equiv\{x,\phi_1\phi_2\}$ --which run in the $x\in [-1,1]$, $\phi_1\in [0,2\pi)$ and $\phi_2\in [0,2\pi)$
domains-- as
\be
\displaystyle
u_1+i\,u_2\,\equiv\,z_1\,=\,\sqrt{\frac{1-x}{2}}e^{i\,\phi_1}\,,
\;\;\;\;\;
u_0+i\,u_3\,\equiv\,z_2\,=\,\sqrt{\frac{1+x}{2}}e^{i\,\phi_2}\,,
\label{coord}
\ee
where we also introduced the two complex coordinates $z_{1,2}$. We can now rewrite the Lagrangian as
\be
L\,=\,-M\,+\,2\lambda\,g_{\alpha\beta} {\dot q}^\alpha{\dot q}^\beta\,,
\label{lag}
\ee
where $g$ is the metric of $S^3$ which reads in our coordinates
\be
ds^2\,=\,g_{\alpha\beta}dq^\alpha dq^\beta\,=\,\frac14\frac1{1-x^2}\,dx^2+\frac{1-x}2\,d{\phi_1}^2+\frac{1+x}2\,d{\phi_2}^2\,.
\ee
The conjugate momenta are $p_{\alpha}=\partial L/\partial {\dot q}^\alpha=4\lambda g_{\alpha\beta}{\dot q}^\beta$
and therefore the classical Hamiltonian is
\be
H_c\,=\,M\,+\,\frac1{8\lambda} p_{\alpha}g^{\alpha\beta}(q)p_{\beta}\,.
\label{cham}
\ee

It should be noted that the points $U$ and $-U$ in what we denoted as the space of collective
coordinates actually describe the same field configuration (see Eq.~(\ref{eq0m},\ref{ansk})).
The $SU(2)=S_3$ manifold we are considering is actually the universal covering of the collective
coordinate space which is given by $S_3/Z_2$. This will be relevant when we will discuss the
quantization.

\subsection{Skyrmion Quantization}

We should now quantize the classical theory described above, by replacing as usual the classical momenta
$p_\alpha$ with the differential operator $-i\partial/\partial q^\alpha$ acting on the wave functions $f(q)$.
Given that the metric depends on $q$, however, there is an ambiguity in how to extract the quantum  hamiltonian
$H_q$ from the classical one in Eq.~(\ref{cham}). This ambiguity is resolved by requiring the quantum theory to
have the same symmetries that the classical one had. At the classical level, we have an $SO(4)\simeq
SU(2)\times SU(2)$
symmetry under $U\rightarrow U\cdot \theta^\dagger$ a
nd $U\rightarrow g\cdot U$ with $\theta,g\in SU(2)$. These correspond,
respectively, to rotations in space and to isospin ({\it i.e.} global vector) transformations, as one can see from
the Ansatz in Eqs.~(\ref{eq0m},\ref{ansk}).
This is because $K$ is invariant under left multiplication by $g$,
and that the Ansatz is left unchanged by performing a rotation
$x_a\sigma^a\rightarrow \theta^\dagger x_a\sigma^a \theta$ and
simultaneously sending $U\rightarrow U\cdot \theta$.
The spin and isospin operators must be given, in the quantum theory, by the generators of these transformations
on the space of wave functions $f(q)$ which are defined by
\be
\left[S^a,U\right]=U\sigma^a/(2)\,,\;\;\;\;\;\left[I^a,U\right]=-\sigma^a/(2)U\,.
\label{comm}
\ee
After a straightforward calculation one finds
\be
\begin{array}{l}
\left\{
\begin{array}{l}
\displaystyle
S^3\,=\,-\frac{i}2\left(\partial_{\phi_1}+\partial_{\phi_2}\right)\\
\displaystyle
S^+\,=\,\frac1{\sqrt{2}}e^{i(\phi_1+\phi_2)}\left[i\sqrt{1-x^2}\partial_x+\frac12\sqrt{\frac{1+x}{1-x}}\partial_{\phi_1}-\frac12\sqrt{\frac{1-x}{1+x}}\partial_{\phi_2}\right]\\
\displaystyle
S^-\,=\,\frac1{\sqrt{2}}e^{-i(\phi_1+\phi_2)}\left[i\sqrt{1-x^2}\partial_x-\frac12\sqrt{\frac{1+x}{1-x}}\partial_{\phi_1}+\frac12\sqrt{\frac{1-x}{1+x}}\partial_{\phi_2}\right]
\end{array}
\right.\\
\left\{
\begin{array}{l}
\displaystyle
I^3\,=\,-\frac{i}2\left(\partial_{\phi_1}-\partial_{\phi_2}\right)\\
\displaystyle
I^+\,=\,-\frac1{\sqrt{2}}e^{i(\phi_1-\phi_2)}\left[i\sqrt{1-x^2}\partial_x+\frac12\sqrt{\frac{1+x}{1-x}}\partial_{\phi_1}+\frac12\sqrt{\frac{1-x}{1+x}}\partial_{\phi_2}\right]\\
\displaystyle
I^-\,=\,-\frac1{\sqrt{2}}e^{-i(\phi_1-\phi_2)}\left[i\sqrt{1-x^2}\partial_x-\frac12\sqrt{\frac{1+x}{1-x}}\partial_{\phi_1}-\frac12\sqrt{\frac{1-x}{1+x}}\partial_{\phi_2}\right]
\end{array}
\right.
\end{array}
\label{spiso}
\ee
where the raising/lowering combinations are $S^{\pm}=(S^1\pm iS^2)/\sqrt{2}$.

The operators in Eq.~(\ref{spiso}) should obey the Hermiticity conditions
$\left(S^3\right)^{\dagger}=S^3$, $\left(S^+\right)^{\dagger}=S^-$, and analogously for the isospin.
In order for the Hermiticity conditions to hold we choose the scalar product to be
\be
\langle A|B\rangle\,\equiv\,\int d^3q\,\sqrt{g}{f_A}^{\dagger}(q)f_B(q)\,,
\label{SC}
\ee
where $\sqrt{g}=1/4$ in our parametrization of $S_3$. The reason why this choice of the scalar product gives the
correct Hermiticity conditions is that $S^a$ and $I^a$ (where $a=1,2,3$) can be written as $X^\alpha\partial_\alpha$
with $X^\alpha$ Killing vectors of the appropriate $S_3$ isometries.
The Killing equation $\nabla_\alpha X_\beta+\nabla_\beta X_\alpha=0$ ensures
the generators to be Hermitian with respect to the scalar product (\ref{SC}).

Knowing that the scalar product must be given by Eq.~(\ref{SC}) greatly helps in guessing what the quantum
Hamiltonian, which has to be Hermitian, should be. We can multiply and divide by $\sqrt{g}$ the kinetic term
of $H_c$ and move one $\sqrt{g}$ factor to the left of $p_\alpha$. Then we apply the quantization rules and
find
\footnote{The last equality holds because $H_q$ is supposed to be acting on the wave functions,
which are scalar functions.}
\be
H_q\,=\,M-\frac1{8\lambda}\frac1{\sqrt{g}}\partial_\alpha \left(\sqrt{g}g^{\alpha\beta}\partial_\beta\right)\,=\,M-\frac1{8\lambda}\nabla_\alpha\nabla^\alpha\,,
\label{qham}
\ee
which is clearly Hermitian. We can immediately show that $H_q$ commutes with spin and isospin, so that the
quantum theory is really symmetric as required: a straightforward calculation gives indeed
\be
H_q\,=\,M+\frac1{2\lambda}S^2\,=\,M+\frac1{2\lambda}I^2\,.
\ee

It would not be difficult to solve the eigenvalue problem for the Hamiltonian (\ref{qham}), but in order to find
the nucleon wave functions it is enough to note that the versor of $n$-dimensional Euclidean space provides the
$n$ representation of the $SO(n)$ isometry group. In our case, $n=4=(2,2)$, which is exactly the
spin/isospin
representation in which nucleons live. It is immediately seen that $z_1$, as defined in Eq.~(\ref{coord}),
has $S^3=I^3=1/2$. Acting with the lowering operators we easily find the wave functions
\be
\displaystyle
\begin{array}{ll}
 |p\,\uparrow\rangle=\displaystyle\frac1{\pi}z_1\,,\;\;\;\;\; & |n\,\uparrow\rangle=\displaystyle\frac{i}{\pi}z_2\,,\\
 |p\,\downarrow\rangle=\displaystyle-\frac{i}{\pi}{\ov z}_2\,,\;\;\;\;\;& |n\,\downarrow\rangle=\displaystyle-\frac{1}{\pi}{\ov z}_1\,,
\end{array}
\ee
which are of course normalized with the scalar product (\ref{SC}). The mass of the nucleons is
therefore $E=M+3/(8\lambda)$.

Notice that the nucleon wave functions are odd under $U\rightarrow -U$, meaning that they are
double-valued on the genuine collective coordinate space $S_3/Z_2$. This corresponds, following
\cite{Finkelstein:1968hy}, to quantize the skyrmion as a fermion and explains how we could get
spin-$1/2$ states after a seemingly bosonic quantization without violating spin-statistic.

Let us now summarize some useful identities which will be used in our calculation. First of all, it is
not hard to check that, after the quantization is performed the rotational velocity becomes
\be
k^a\,=\,-i\,{\rm Tr}\left[U^{\dagger}{\dot U}\sigma^a\right]\,=\,\frac1\lambda S^a\,,
\ee
and analogously
\be
i\,{\rm Tr}\left[{\dot U}U^{\dagger}\sigma^a\right]\,=\,\frac1\lambda I^a\,.
\label{Iden}
\ee
Other identities which we will use in our calculations are
\bea
\displaystyle
&&\langle {\rm Tr}\left[U\,\sigma^b U^\dagger\sigma^a\right] = -\frac83 S^b I^a\rangle\,, \nn\\
&&\langle {\rm Tr}\left[U\,\sigma^b \x_b (k\cdot\x) U^\dagger\sigma^a\right] = -\frac2{3\lambda} I^a\rangle\,,
\label{qr}
\eea
where the VEV symbols $\langle...\rangle$ mean that those are not operatorial identities, but they only hold when
the operators act on the subspace of nucleon states. Notice that the second equation in (\ref{qr}) is implied by
the first one if one also uses the commutation relation (\ref{comm}), Eq.~(\ref{Iden}) and the fact that, on
nucleon states, $\langle\left\{S^a,S^i\right\}=\delta^{ai}/2\rangle$.

\subsection{The Nucleon Form Factors}

The nucleon form factors parametrize the matrix element of the currents on two
nucleon states. For the isoscalar and isovector currents we have
\bea\label{eqEMCurrCorr}
\displaystyle
\langle N_f(p') | J^\mu_{S}(0) | N_i(p)\rangle = \bar u_f(p') \left[
F_1^S(q^2) \gamma^\mu + \frac{i F_2^S(q^2)}{2 M_N} \sigma^{\mu\nu}q_\nu\right]u_i(p),\nn\\
\langle N_f(p') | J^{\mu a}_{V}(0) | N_i(p)\rangle = \bar u_f(p') \left[
F_1^V(q^2) \gamma^\mu + \frac{i F_2^V(q^2)}{2 M_N} \sigma^{\mu\nu}q_\nu\right]\left(2I^a\right) u_i(p),
\eea
where the currents are
defined as $J_{V}^a=J_R^a+J_L^a$ and $J_S=1/3\left({\widehat J}_R+{\widehat J}_L\right)$
in terms of the chiral ones. In the equation above
$q \equiv p'-p$ is the $4$-momentum transfer, $N_i$ and $N_f$ are the initial and final
nucleon states and $u_i(p)$, $\bar u_f(p')$ their wave functions,
$I^a=\sigma^a/2$ is the isospin generators and
$\sigma^{\mu\nu} \equiv i/2 [\gamma^\mu, \gamma^\nu]$.
For the axial current $J_{A}^a=J_R^a-J_L^a$ we have
\be\label{eqAxCurrCorr}
\displaystyle
\langle N_f(p') | J^a_{A \mu}(0) | N_i(p)\rangle = \bar u_f(p')
G_A(q^2) \left[ \gamma_\mu -\frac{2 M_N}{q^2} q^\mu \right]\gamma^5 I^a u_f(p)\,.
\ee
Exact axial and isospin symmetries, which hold in our model, have been assumed in the
definitions above.

In our non-relativistic model the current correlators will be computed in the Breit frame
in which the initial nucleon
has 3-momentum $-{\vec q}/2$ and the final $+{\vec q}/2$ (i.e. $p^\mu = (E, -{\vec q}/2)$ and
$p'^\mu = (E, {\vec q}/2)$, and $q^2=-{\vec q\,}^2$, with $E=\sqrt{M_N^2 + {\vec q\,}^2/4}$).
Notice that the textbook definitions in Eqs.~(\ref{eqEMCurrCorr},\ref{eqAxCurrCorr})
involve nucleon states which are normalized with $\sqrt{2 E}$; in order to match with
our non-relativistic normalization we have to divide all correlators by $2M_N$. The
vector currents become
\bea
\displaystyle
\langle N_f({\vec q}/2) | J^0_{S}(0) | N_i(-{\vec q}/2)\rangle &=&  G_{E}^{S}({\vec q\,}^2) \chi_f^\dagger \chi_i\,,\nn\\
\displaystyle
\langle N_f({\vec q}/2) | J^i_{S}(0) | N_i(-{\vec q}/2)\rangle &=& i\, \frac{G_M^{S}({\vec q\,}^2)}{2 M_N} \chi_f^\dagger 2 ({\vec S} \times {\vec q})^i \chi_i\,,\nn\\
\displaystyle
\langle N_f({\vec q}/2) | J^{0 a}_{V}(0) | N_i(-{\vec q}/2)\rangle &=& G_{E}^{V}({\vec q\,}^2) \chi_f^\dagger \left(2I^a\right) \chi_i\,,\nn\\
\displaystyle
\langle N_f({\vec q}/2) | J^{i a}_{V}(0) | N_i(-{\vec q}/2)\rangle &=& i\, \frac{G_M^{V}({\vec q\,}^2) }{2 M_N}
\chi_f^\dagger 2 ({\vec S}  \times {\vec q})^i \left(2I^a\right) \chi_i\,,
\label{bf}
\eea
where we defined
\be
G_E^{S,V}(-q^2) = F_1^{S,V}(q^2) + \frac{q^2}{4 M_N^2} F_2^{S,V}(q^2)\,, \qquad G_M^{S,V}(-q^2) = F_1^{S,V}(q^2) + F_2^{S,V}(q^2)\,,
\ee
and used the definition
$({\vec S} \times {\vec q})^i \equiv \varepsilon^{ijk} S^j q^k$.
The nucleon spin/isospin vectors of state $\chi_{i,f}$  are
normalized to $\chi^\dagger \chi = 1$.
For the axial current we find
\bea
\langle N_f({\vec q}/2) | J_A^{i,a}(0) | N_i(-{\vec q\,}/2)\rangle &=&
\chi_f^\dagger \frac{E}{M_N} G_A({\vec q\,}^2) 2 S^i_T
\frac{\tau^a}{2}\chi_i\,,\nn\\
\langle N_f({\vec q\,}/2) | J_A^{0,a}(0) | N_i(-{\vec q\,}/2)\rangle &=&0
\label{bfa}
\eea
where ${\vec S}_T \equiv {\vec S} - \hat{{\vec{q}}}\  {\vec S}\cdot{\hat{\vec{q}}}$ is the transverse  component of the spin operator.

It is straightforward to compute the matrix elements of the currents in position space on
static nucleon states. Plugging the Ansatz (\ref{sts},\ref{eq0m},\ref{ansk1},\ref{ansk}) in
the definition of the currents (\ref{cur0}) and performing the quantization one obtains
quantum mechanical operators acting on the nucleons. The matrix elements are easily
computed using the results of sect.~3.1. We finally obtain the form factors by taking the
Fourier transform and comparing with Eqs.~(\ref{bf},\ref{bfa}). We have
\footnote{It is quite intuitive that the form factors can be computed in this way.
Given that solitons are infinitely heavy at small coupling, in the Breit frame
they are almost static during the process of scattering with the current.
To check this, however, we should perform the
quantization of the collective coordinates associated with the center-of-mass motion, as it
was done in \cite{Braaten:1986iw} for the original 4D Skyrme model.}
\bea
\displaystyle
&&G_E^S\,=\,-\frac{N_c}{6\pi\gamma L}\int dr\,r\,j_0(qr)\left(a(z)\partial_zs\right)_{UV}\nn\\
&&G_E^V\,=\,\frac{4\pi M_5}{3\lambda}\int dr\,r^2\,j_0(qr)\left[a(z)\left(\partial_zv-2\left(D_z\chi\right)_{(2)}\right)\right]_{UV}\nn\\
&&G_M^S\,=\,\frac{8\pi M_NM_5\alpha}{3\lambda}\int dr\,r^3\,\frac{j_1(qr)}{qr}\left(a(z)\partial_zQ\right)_{UV}\nn\\
&&G_M^V\,=\,\frac{M_N\,N_c}{3\pi L\gamma \alpha}\int dr\,r^2\,\frac{j_1(qr)}{qr}\left(a(z)\left(D_z\phi\right)_{(2)}\right)_{UV}\nn\\
&&G_A\,=\,%\frac{M_N}{E}
\frac{N_c}{3\pi\alpha\gamma L}
\int dr\,r\left[
a(z)\frac{j_1(qr)}{qr}
\left(\left(D_z\phi\right)_{(1)}-r\,A_{zr}\right)
-a(z)\left(D_z\phi\right)_{(1)} j_0(qr)
\right]_{UV}
\label{cff}
\eea
where $j_n$ are spherical Bessel functions which arise because of the Fourier transform.

\section{Properties of baryons: Results}
\label{results}
%AG Sect.~4

In this section we will present our results. After discussing some qualitative features,
such as the large-$N_c$ scaling of the form factors and the divergences of the isovector radii due to exact
chiral symmetry, we extrapolate to the physically relevant case of $N_c=3$ and perform a quantitative
comparison with the experimental data. Consistently with our working hypothesis that the 5D model really
describes large-$N_c$ QCD we find a $30\%$ relative discrepancy.

\subsubsection*{Large-$N_c$ Scaling}

As explained in sect.~2.1, all the three parameters $\alpha$, $\gamma$ and $L$ of our 5D model
should scale like $N_{c}^0$, Eq.~(\ref{scaling}),  in order for the large-$N_c$ scaling of meson couplings and  masses to be correctly reproduced.
% Eq.~(\ref{gamma}) therefore implies that the coupling $M_5$ grows like $N_c$ and the semiclassical
%expansion in 5D coincides with the $1/N_c$ expansion on the 4D side. 
%Notice that these scaling of the parameters
%are uniquely dictated by what we know to be 
This implies the following scaling for the baryon observables.
First, we notice that  the solitonic solution is independent of $N_c$ given 
that $M_5$ factorizes out of the action and does not appear in the EOM. 
This implies that the radii of the soliton does not scale with  $N_c$, while
the classical mass $M$ and the moment of inertia $\lambda$  scale like $N_c$.
Using this we can read the $N_c$-scaling of the electric and magnetic form factors
%, obtained from the matrix elements  Eq.~(\ref{bf}), 
%seems naively  to scale like $N_c$ and $M_N N_c$ respectively 
%since the currents Eq.~(\ref{cur0}) contains a factor $M_5$.
%Nevertheless,  we find, due to cancellations, deviations from this naive scaling for 
%$G_{E}^V$ and  $G_{M}^S$,  as can be seen  from 
from Eq.~(\ref{cff}):
\begin{equation}
\label{scaba}
G_E^S\sim N_c\ ,\ \ \ G_E^V\sim N_c^0\ , \ \ \ \frac{G_M^S}{M_N}\sim N_c^0\ , \ \ \  \frac{G_M^V}{M_N}\sim N_c\, .
\end{equation}

In large-$N_c$ QCD the baryon masses scale like $N_c$  \cite{Witten:1979kh}, as in our model. 
The matrix elements of the currents on nucleon states are also expected 
 to scale like $N_c$, even though cancellations are  possible \cite{Manohar:1998xv}.
The radii, therefore, must scale like $N_c^0$ as we find and, looking 
at the definition (\ref{bf}), $G_E^{S,V}$ and $G_M^{S,V}/M_N$ should both scale 
like $N_c$ up to cancellations. It is very simple to understand why, both in QCD and 
in our model, there must be a cancellation in $G_{E}^V$.
Remembering that the temporal component
of the current at zero momentum gives the conserved charge and looking at the definitions (\ref{bf}),
 one immediately
obtains  $G_{E}^V(0)=1/2$ because 
the skyrmion,  as the nucleon, is in the $1/2$ representation of isospin.
This condition is respected by our model as  it is  implied
by the EOM, and  fulfilled to great accuracy ($0.1\%$) by the numerical solution.
Similarly we find at zero momentum $G_{E}^S(0)=N_c/6$  as required for a bound-state  
 made of  $N_c$ quarks  of $U(1)_V$ charge  $1/6$ each (in our conventions). Also this 
condition is implied by the EOM and verified by the numerical solution.

%The reason for $G_{E}^V\sim N_c^0$  is simple to understand. 
%Remembering that the temporal component
%of the current at zero momentum gives the conserved charge and looking at the definitions (\ref{bf}), one immediately
%obtains  $G_{E}^V(0)=1/2$, because 
%the skyrmion,  as the nucleon, is in the $1/2$ representation of isospin.
%This condition is respected by our model as  it is  implied
%by the EOM, and  fulfilled to great accuracy ($0.1\%$) by the numerical solution.
%Similarly we find at zero momentum $G_{E}^S(0)=N_c/6$  as required for a bound-state  
% made of  $N_c$ quarks  of $U(1)_V$ charge  $1/6$ each (in our conventions).
%
%All these $N_c$-scalings of the physical quantities of our 5D soliton 
%are in agreement with those expected for baryons in  large-$N_c$ QCD \cite{Witten:1979kh}. 
%The masses of the baryons are known to grow with $N_c$, while  their radii scale with $N_c^0$.
%Also the scaling  Eq.~(\ref{scaba}) can be deduced   in  large-$N_c$ QCD,
%except $G_{M}^S/M_N\sim N_c^0$  that we have not been able to ... {\it find any reason?}
Concerning the second cancellation, {\it i.e.} $G_{M}^S/M_N\sim N_c^0$, we are not 
able to prove that it must take place in large-$N_c$ QCD as it does in our model. 
We can, however,
check that it occurs in the naive quark model, or better in its generalization for arbitrary odd $N_c=2\,k+1$
\cite{Witten:1983tx}. In this non-relativistic model the Nucleon wave function is made of $2k+1$ quark states $q_i$, $2k$
of which are collected into $k$ bilinear spin/isospin singlets while the last one has free indices which give to the
Nucleon its spin/isospin quantum numbers. Of course, the wave function
is symmetrized in flavor and spin given that the color indices are contracted with the antisymmetric tensor and the
spatial wave function is assumed to be symmetric. The current operator is the sum of the currents for the $2k+1$
quarks, each of which will assume by symmetry the same form as in Eq.~(\ref{bf}). If $S_{1,2}$ and $I_{1,2}$
represent the spin and isospin operators on the quarks $q_{1,2}$ the
operators $S_1+S_2$ and $I_1+I_2$ will vanish on the
singlet combination of the two quarks, but $S_1I_1+S_2I_2$ will not.
The $k$ singlets will therefore only contribute to $G_{E}^S$, $G_{M}^V$ and $G_{A}$, which will have the
naive scaling, while for the others we find cancellations.

A detailed calculation can be found in \cite{Karl:1984cz} where, among other things, the proton and neutron
magnetic moments and the axial coupling are computed in the naive quark model. The magnetic moments are related
to the form factor at zero momentum as $\mu_V/\mu_N=G_{M}^V(0)$ and $\mu_S/\mu_N=G_{M}^S(0)$ where $\mu_N=1/(2M_N)$
is the nuclear magneton and $2\,\mu_V=\mu_p-\mu_n$, $2\,\mu_S=\mu_p+\mu_n$. In accordance with the previous discussion,
the results in the naive quark model are $2\mu_S=\mu_u+\mu_d$ and $2\mu_V=2k/3(\mu_u-\mu_d)$, where $\mu_{u,d}$
are the quark magnetic moments, while for the axial coupling one finds $g_A=G_A(0)=2k/3+1$ which scales like $N_c$
as expected:
\begin{equation}
g_A=\frac{N_c}{3}+\frac{2}{3}\, .
\end{equation}
Notice that for $N_c=3$ the subleading term in the  $1/N_c$-expansion  represents a   $60\%$
correction.
% Notice that the $1/N_c$ corrections  to the axial coupling are quite big in
%the naive quark model: for $k=1$ ($N_c=3$) the leading term contributes as $1$ while the 
%full  results is $5/3$, which is $60\%$ bigger. 
We have of course no reason to believe that such big corrections should
persist in the true large-$N_c$  QCD; this remark simply suggests that
``large'' $1/N_c$ corrections to the form factors are not excluded.

\subsubsection*{Divergences in the Chiral Limit}

It is well known that in QCD the isovector electric $\langle r_{E,\,V}^2\rangle$ and magnetic
$\langle r_{M,\,V}^2\rangle$ radii which are proportional, respectively, to the $q^2$ derivative of $G_{E}^V$
and $G_{M}^V$ at zero momentum, diverge in the chiral limit \cite{Beg:1973sc}.
In our model, as in the Skyrme model, divergences in the integrals of Eq.~(\ref{cff}) which define the form
factors are due, as in QCD, to the massless pions. If all the fields were massive, indeed, any solution to the
EOM would fall down exponentially at large $r$ while in the present case power-like behaviors can appear. These
power-like terms in the large-$r$ expansion of the solution can be derived analytically by performing a Taylor
expansion of the fields around infinity ($1/r=0$), substituting into the EOM and solving order by order in $1/r$.
%For each equation we stop to a suitable $1/r$ order such that the leading $1/r$ correction is switched on for all
%the fields.
The exponentially suppressed part of the solution will never contribute to the expansion. 
%In the gauge in
%which (the form factors are, of course, gauge invariant) the topological twist is at the origin $r=0$ and the solution
%is trivial for $r\rightarrow \infty$ the first few terms are \footnote{In the equations which follow we put $L=1$ for simplicity.}
%\be
%\left\{
%\begin{array}{l}
%\displaystyle A_1 = \frac{2 z (z-1)}{r^3} \beta\\
%\displaystyle A_2 = \frac{\beta}{r^2} + \frac{4 z^3 - 6 z^2 +1}{2 r^4} \beta\\
%\displaystyle \phi_1 = \frac{z (1-z)}{r^2} \beta + \frac{z(z^3 - 2 z^2 +1)}{2 r^4}\beta\\
%\displaystyle \phi_2 = -1 + \frac{z^2 (3-2z)}{2 r^4} \beta^2\\
%\displaystyle s = \frac{z^2(z^6 - 4 z^4 +8)}{4 r^8} \gamma \beta^3
%\end{array}
%\right.
%\qquad
%\left\{
%\begin{array}{l}
%\displaystyle \chi_1 = \frac{z (z-1)}{r^2} \beta\\
%\displaystyle \chi_2 = 1 + \frac{z^2 (2 z-3)}{2 r^4} \beta^2\\
%\displaystyle v = -1 + \frac{z^2 (z^2-3)^2}{12 r^6} \beta^2\\
%\displaystyle q = - \frac{z^2 (z^6 - 4 z^4+8)}{4 r^8} \gamma \beta^3
%\end{array}
%\right.
%\label{larger}
%\ee
%where $\beta$ is an unknown parameter which depends on the entire solution and can only be determined
%numerically. We checked that the large-$r$ behavior of our numerical solution is very well
%approximated by Eq.~(\ref{larger}).
This procedure allows us to determine the asymptotic expansion of the solution completely, 
up to an integration constant $\beta$.
Substituting the expansion into the definitions of the form factors (\ref{cff})
one gets
\be
\left\{\displaystyle
\begin{array}{l}
\displaystyle G_E^S\;\propto\; \displaystyle\beta^3 \int dr \frac{1}{r^7} j_0(q r)+\ldots\\
\displaystyle G_E^V \;\propto\; \displaystyle\beta^2 \int dr \frac{1}{r^2} j_0(qr)+\ldots\\
 G_M^S\;\propto\; \displaystyle\beta^3 \int dr \frac{1}{r^5}  \frac{j_1(qr)}{qr}+\ldots\\
G_M^V\;\propto\; \displaystyle\beta^2 \int dr \frac{1}{r^2}   \frac{j_1(qr)}{qr}+\ldots
\end{array}
\right.\,.
\label{eq:largerbehaviour}
\ee
All the form factors are finite for any $q$, including $q=0$.
The electric and magnetic radii, however, are defined as
\be
\langle r^2_{E,M} \rangle = -\frac{6}{G_{E,M}(\vec q\,^2 = 0)}
\left.\frac{d G_{E,M}(\vec q\,^2)}{d \vec q\,^2}\right|_{\vec q\,^2=0}\,,
\label{eq:radii}
\ee
and taking a $q^2$ derivative of Eqs.~(\ref{eq:largerbehaviour}) makes one more power of $r^2$ appear
in the integral. It is easy to see that the scalar radii are finite, while the vector ones are divergent.
%as anticipated. We will now discuss the axial coupling and the axial radius and show that both are finite.
%
%The expression in Eq.~(\ref{cff}) for the axial form factor $G_A$
%presents some subtleties for vanishing $q^2$.
%Given the asymptotic expansion of the solution in Eq.~(\ref{larger}) the
%axial coupling integral behaves for large $r$ like
For the axial form factor $G_A$ we find
\be
\displaystyle G_A\;\propto\;
\int dr\left[
 \left(\frac{3}{r} \beta - \frac{1}{r^5} \beta^3\right)\frac{j_1(qr)}{qr}
+ \left(-\frac{1}{r} \beta + \frac{5}{7 r^5} \beta^3\right) j_0(qr)+\ldots\right]\,,
\label{axas}
\ee
%where the leading $1/r$ terms (the ones which are linear in $\beta$) can be obtained from
%Eq.~(\ref{larger}) while for the others one needs higher order terms which are not reported
%in Eq.~(\ref{larger}). 
The integral in Eq.~(\ref{axas}) is convergent for any $q\neq0$ but, however,
it is not uniformely convergent for $q \rightarrow 0$.
%, however, the integral
%is not uniformly convergent and one cannot exchange the limit with the integration. 
The leading
$1/r$ term in Eq.~(\ref{axas}) is indeed given by
$I(q) = \beta\int_0^\infty dr\ (1/r)\left(3 j_1(qr)/(qr)-j_0(qr)\right)$, 
which is independent of $q$
and equal to $\beta/3$, while the argument of the integral vanishes for $q\rightarrow0$
so that 
exchanging the limit and integral operations would give the wrong result $I(0)=0$.
To restore uniform convergence and obtain an analytic formula
for $g_A$ one can subtract the $I(q)$ term from the expression
in Eq.~(\ref{cff}) for $G_A$ and replace it with $\beta/3$.
Rewriting the axial form factor in this way is also useful to establish that
the axial radius, which seems divergent if looking at Eq.~(\ref{axas}), is
on the contrary finite. The $I(q)$ term, indeed, does not contribute
to the $q^2$ derivative and the ones which are left in Eq.~(\ref{axas}) give
a finite contribution.

We have found, compatibly with the QCD expectation, that all the form factors and radii are 
finite but the isovector ones. Notice that the structure of the divergences is completely 
determined by the asymptotic large-$r$ behaviour of the solution, and not by its detailed 
form ({\it i.e.}, for instance, by the actual value of the integration constant $\beta$ which
depends on the entire solution). 
Our model coincides, in the IR, with the Skyrme model, therefore 
the asymptotic behaviour of the current densities is expected to be the same in the two cases.
%This is indeed what we find, and 
This explains why we obtained the same divergences as in the Skyrme model.

\subsubsection*{Pion Form Factor and Goldberger-Treiman relation}

It is of some interest to define and compute the pion-nucleon form factor which parametrizes the matrix element
on Nucleon states of the pion field. In the Breit frame (for normalized nucleon states)
 it is
\be
\displaystyle
\langle N_f({\vec q}/2) | \pi^{a}(0) | N_i(-{\vec q\,}/2)\rangle =
-\frac{i}{2M_N{\vec q}^2}G_{NN\pi}({\vec q}^2)\chi_f^\dagger(2 S^i) q_i (2 I^a)\chi_i\,,
\label{dme}
\ee
where $\pi^a(x)$ is the normalized and ``canonical" pion field operator. The field is canonical
in the sense that its quadratic effective lagrangian only contains the canonical kinetic term
${\mathcal L}_2=1/2(\partial\pi_a)^2$, or equivalently that its propagator is the canonical one,
without a non-trivial form factor. With this definition, $G_{NN\pi}$ is the vertex form factor of the
meson-exchange model for nucleon-nucleon interactions \cite{Machleidt:1987hj} and corresponds
to an interaction \footnote{Nucleon scattering, in our model, is a soliton scattering process and we
have no reason to believe that it can be described by meson-exchange, {\it i.e.} that contact terms
are suppressed. Therefore, we will not attempt any comparison of our form factor with the one
used in meson-exchange models.}
\be
{\mathcal L}_{NN\pi}=i\, (G_{NN\pi}(\Box)\pi_a){\ov N}\gamma^\mu\gamma_5(2I^a)N\,.
\ee
On-shell, the form factor reduces to the pion-nucleon coupling constant, $G_{NN\pi}(0)=g_{NN\pi}$,
whose experimental value is $g_{NN\pi}=13.5\pm0.1$.

The pion field which matches the requirements above is given by the zero-mode of the
KK decomposition. In the unitary gauge $\partial_z(a(z) A_5) = 0$,
where $A_M \equiv (L_M-R_M)/2$, and for AdS$_5$ space,  one has
\be
\displaystyle A^{(un)}_5(x, z) = \frac{1}{F_\pi L}\frac1{a(z)} \pi^a(x)\sigma_a\, ,
\ee
where $F_\pi$ is given in Eq.~(\ref{fpi}).
Gauge-transforming back to the gauge in which our numerical solution is provided and using
the Ansatz in Eqs.~(\ref{sts},\ref{eq0m}) we find the pion field
\be
\displaystyle
\pi^a\,=\,-\frac{F_\pi}{2} \int_{z_{\UV}}^{z_{\IR}} dz A_2(r,z){\widehat x}^b
{\rm Tr}\left[U\sigma_bU^\dagger \sigma^a\right]\,.
\ee
Taking the matrix element of the above expression and comparing with Eq.~(\ref{dme})
one obtains
\be
G_{NN\pi} (q^2) =  -\frac{8 \pi}{3} M_N F_\pi q
\int_{0}^\infty dr j_1(q r) \int d z\, r^2\,A_2(r,z)\,.
\label{rpc}
\ee
At $q\rightarrow0$  the form factor  $G_{NN\pi}$ is completely determined by the large-$r$ behavior of the field $A_2$, given by $A_2\rightarrow \beta/r^2$. 
%Due to the $q$ factor, indeed, only the divergent part of the integral contributes. 
We then find
\be
\label{gnnp}
g_{NN\pi} = -\frac{32\pi }{3 } M_NF_\pi{\beta L^2}\, .
\ee
By using Eqs.~(33,34) of Ref.~\cite{Adkins:1983ya} which show that also $g_A$ is determined by the asymptotic behavior of the axial current, one finds 
\be
g_A = -\frac{32\pi}{3} F^2_\pi\beta L^2\, ,
\ee
 that, together with Eq.~(\ref{gnnp}),  leads to the famous Goldberger-Treiman relation $F_\pi g_{\pi NN}=M_N g_A$.
This relation, which is a consequence of having exact chiral symmetry,
 has been numerically verified  to $0.01\%$.

\subsubsection*{Comparison with Experiments}

\begin{table}
\tbl{Prediction of the nucleon observables with the microscopic
parameters fixed by a fit on the mesonic observables.
The deviation from the empirical data
is computed using the expression $(th-exp)/\min(|th|, |exp|)$, where
$th$ and $exp$ denote, respectively, the prediction of our model and the
experimental result.}
%\centering{
{\begin{tabular}{c@{\hspace{2em}}c@{\hspace{2em}}c@{\hspace{2em}}c}
\hline
& Experiment & AdS$_5$ & Deviation\\
\hline
$M_N$ & $940\ \text{MeV}$ & $1130\ \text{MeV}$ & $+20\%$\\
$\mu_S$ & $0.44$ & $0.34$ & $-30\%$\\
$\mu_V$ & $2.35$ & $1.79$ & $-31\%$\\
$g_A$ & $1.25$ & $0.70$ & $-79\%$\\
$\sqrt{\langle r_{E,S}^2\rangle}$ & $0.79\ \text{fm}$ & $0.88\ \text{fm}$ & $+11\%$\\
$\sqrt{\langle r_{E,V}^2\rangle}$ & $0.93\ \text{fm}$ & $\infty$ \\
$\sqrt{\langle r_{M,S}^2\rangle}$ & $0.82\ \text{fm}$ & $0.92\ \text{fm}$ & $+12\%$\\
$\sqrt{\langle r_{M,V}^2\rangle}$ & $0.87\ \text{fm}$ & $\infty$ \\
$\sqrt{\langle r_{A}^2\rangle}$ & $0.68\ \text{fm}$ & $0.76\ \text{fm}$ & $+12\%$\\
\hline
$\mu_p/\mu_n$ & $-1.461$ & $-1.459$ & $+0.1\%$\\
\hline
\hline
\end{tabular}}
\label{tab:MesonFit}
\end{table}

Let us  now compare our results with real-world QCD.  
We  therefore fix
the number of colors $N_c=3$ and choose our microscopic parameters
to be those  that gave the best fit to the mesonic quantities:
 $1/L \simeq 343\ \text{MeV}$,
$M_5 L \simeq 0.0165$ and $\alpha \simeq 0.94$ ($\gamma \simeq 1.23$).
The numerical results of our analysis and the deviation with respect to the
experimental data are reported in table~\ref{tab:MesonFit}.
We find a fair agreement with the experiments, a $36\%$ total RMSE
which is compatible with the expected size of $1/N_c$ corrections.
%We discussed in the previous section that
%the isovector radii are divergent because of the chiral limit,
%it would be interesting to add the pion mass to the model and compute these observables.
The axial charge $g_A$ is the one which shows
the larger ($80\%$) deviation, and indeed removing this observable the RMSE decreases to $21\%$. 
We cannot exclude that, in a theory in which the naive expansion parameter is $1/3$,
enhanced $80\%$  corrections to few observables might appear at the next-to-leading order.
Nevertheless,  we think that this result could be very sensitive to the pion mass
and therefore could be substantially improved  
in  5D models  that incorporate  explicit chiral breaking.
The reason for this is that $g_A$  is strongly sensitive to the large-$r$ behavior of
the solution (see the discussion following Eq.~(\ref{axas}))
which is in turn heavily affected by the presence of the pion mass.
Notice that a larger value,  $g_A\simeq 0.99$, is obtained 
in the ``complete'' model described in Ref.~\cite{Meissner:1987ge}, a model
with similar features to our 5D scenario and which includes a nonzero pion mass.
This expectation, however, fails in the original Skyrme model,
where the addition of the pion mass does not affect $g_A$ significantly \cite{Adkins:1983hy}
and one finds  $g_A\simeq 0.65$.

Table~\ref{tab:MesonFit} also shows the proton-neutron magnetic moment ratio, $\mu_p/\mu_n$, which is
in perfect agreement with the experimental value. 
This observable  is the only one in the list that 
includes two orders of the $1/N_c$ expansion. Indeed,
due to the  scaling  $\mu_V\sim N_c$ and $\mu_S\sim N_c^0$, we have 
$\mu_p/\mu_n=-(\mu_V+\mu_S)/(\mu_V-\mu_S)\simeq -1-2\mu_S/\mu_V$.

\begin{figure}[t]
\centerline{ \hspace*{-0.5cm}
\includegraphics[width=0.46 \textwidth]{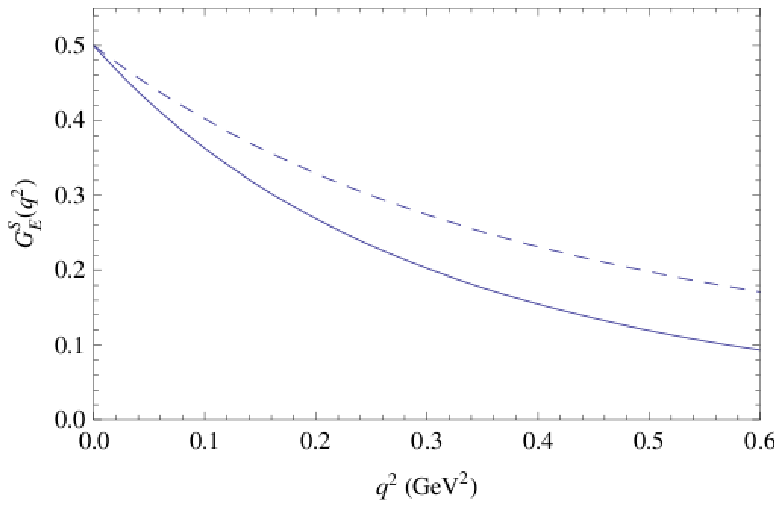}
\hspace*{0.5cm}
\includegraphics[width=0.46 \textwidth]{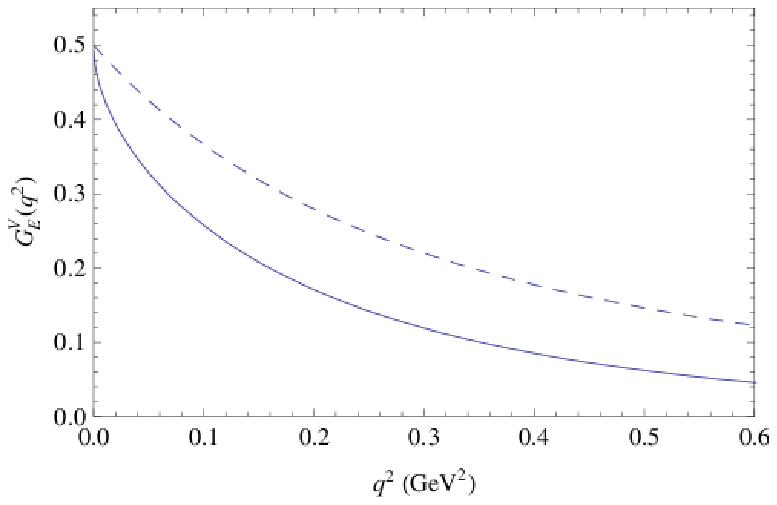}
}
\caption{Scalar (left) and vector (right) electric form factors.
We compare the results with the empirical dipole fit (dashed line) \cite{Meissner:1987ge}.}
\label{Fig:ElectricFormFactors}
\end{figure}

\begin{figure}[t]
\centerline{ \hspace*{-0.5cm}
\includegraphics[width=0.46 \textwidth]{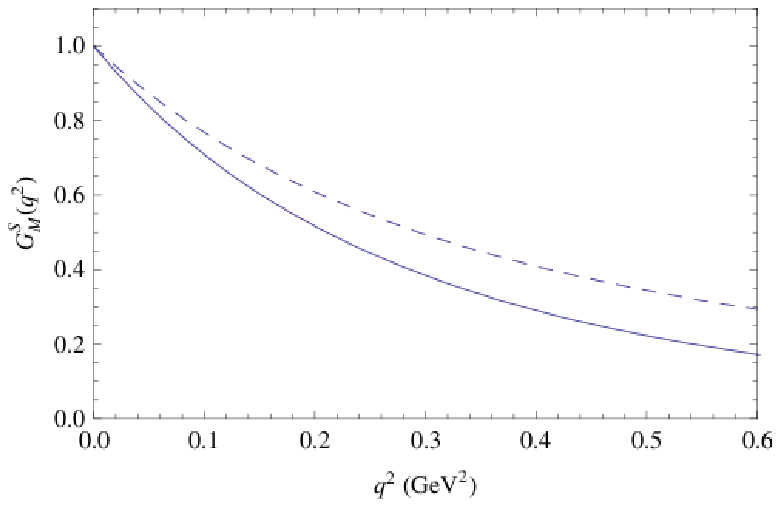}
\hspace*{0.5cm}
\includegraphics[width=0.46 \textwidth]{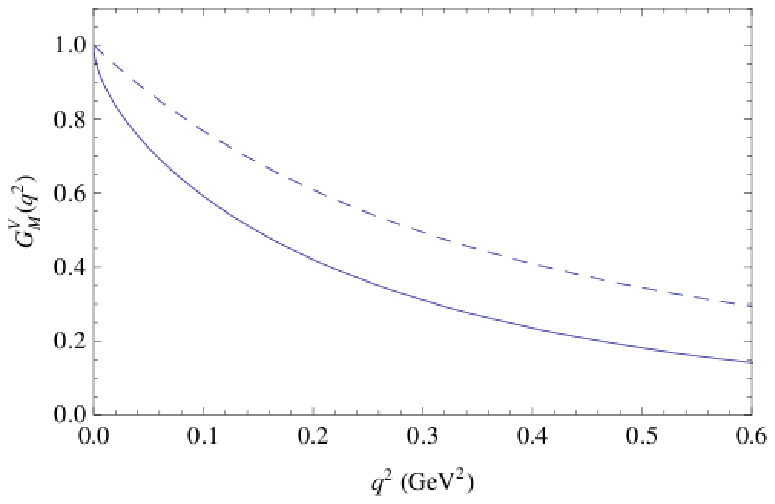}
}
\caption{Normalized scalar (left) and vector (right) magnetic form factors.
We compare the results with the empirical dipole fit (dashed line) \cite{Meissner:1987ge}.}
\label{Fig:MagneticFormFactors}
\end{figure}

\begin{figure}[t]
\centerline{
\includegraphics[width=0.46 \textwidth]{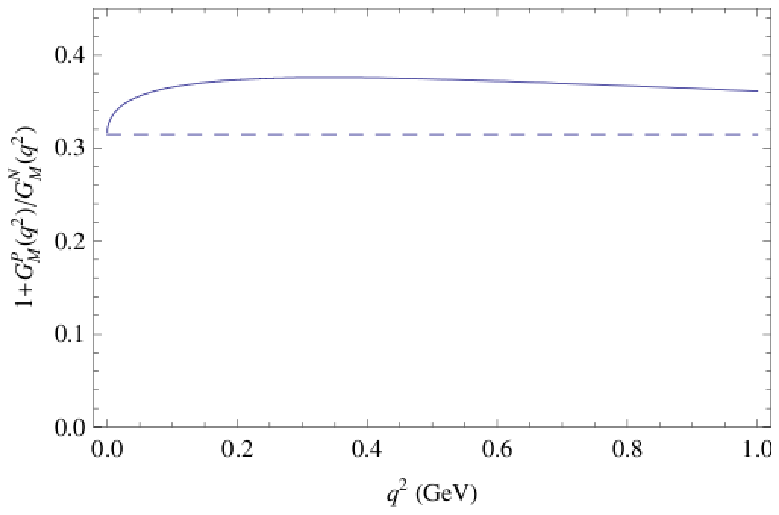}
\hspace*{0.5cm}
\includegraphics[width=0.46 \textwidth]{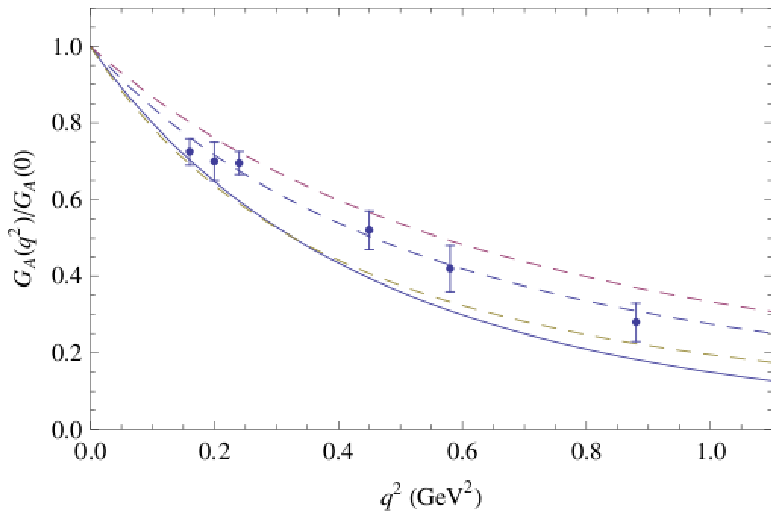}
}
\caption{Left: deviation of the ratio of proton and neutron magnetic form factors
from the large $N_c$ value (solid line), compared with the dipole fit
of the experimental data (dashed line).
Right: normalized axial form factor (solid line) compared with
the empirical dipole fit
(dashed lines) \cite{Meissner:1987ge} and with
the experimental data taken from \cite{Amaldi:1972vf, DelGuerra:1976uj}.}
\label{Fig:AxialFormFactor}
\end{figure}

In figs.~\ref{Fig:ElectricFormFactors}, \ref{Fig:MagneticFormFactors}
and \ref{Fig:AxialFormFactor} we compare the normalized nucleon form factors
at $q^2 \neq 0$ with the dipole fit of the experimental data.
The shape of the scalar and axial form factors
is of the dipole type, the discrepancy is mainly due to the error in the radii.
The shape of vector form factors is of course not of the dipole type for small
$q^2$, but this is due to the divergence of the derivative at $q^2=0$.
Including the pion mass will for sure improve the situation given that
it will render finite the slope at zero momentum; it would be interesting
to see if the dipole shape of these form factors is recovered in the presence of the pion mass.
 We also plot in the left panel of fig.~\ref{Fig:AxialFormFactor} the deviation of
ratio of the proton and neutron magnetic form factors from the large $N_c$
value which is given, due to the the different
large-$N_c$ scaling of the isoscalar and isovector components, by
$G_M^P(q)/G_M^N(q) = -1$. Not only
we find that this quantity is quite well predicted, with an error $\lesssim 15\%$,
but we also see that its shape, in agreement with observations, is nearly constant
away from $q^2=0$. Also in this case corrections from the pion mass are expected to go
in the right direction.

%\section{Baryons in other 5D models}

%Sect.~2 of AA1(Alex-Andrea-1)

\section{Conclusions and outlook}

We have shown that  five-dimensional models, used to describe  meson properties of  QCD, 
can also be considered to study baryon physics.
Baryons appear in these theories as soliton  of  sizes of order  $1/m_\rho$ stabilized
by the presence of the CS term.
We have reviewed the procedure to calculate the static properties of the nucleons
that have shown to be in  reasonable good agreement with the experimental data. 
This shows, once again, that   5D models provide an alternative and very  promising   tool to study  properties of QCD in certain regimes.

There are  further  issues that deserve  to be analyzed. 
The most urgent one is the inclusion of a nonzero pion mass. 
As we have pointed out above,  this will be crucial
to calculate the isovector radii and, maybe, improve the prediction for $g_A$.
For this purpose we need to use a 5D model along the lines  of Ref.~\cite{Erlich:2005qh,Rold:2005} where an explicit breaking of the chiral symmetry, corresponding to the quark masses,
is introduced.
We can also use this approach to study  systems with  high baryon densities,  
analyze possible phase transitions or study the properties of nuclear matter.

\section*{Acknowledgments}

The  work  of AP  was  partly supported   by the
Research Projects CICYT-FEDER-FPA2005-02211,
SGR2005-00916 and   "UniverseNet" (MRTN-CT-2006-035863).
AW thanks G.~Panico for the many useful discussions.

\appendix

\section{Numerical Methods}

In this technical appendix we explain how the numerical determination of the 
soliton solution is performed.

\subsection*{Equations of Motion and Boundary Conditions}

Let us first of all write down the EOM for the 2D fields which characterize 
our Ansatz in Eqs.~(\ref{sts}), (\ref{ansk}), (\ref{eq0m}) and (\ref{ansk1}). These 
can be obtained by plugging the Ansatz either directly in the 5D equations (\ref{eomt}) 
or in the 5D action in Eqs.~(\ref{Sg}) and (\ref{Scs}). In the second case one 
gets the 2D action specified by Eqs.~(\ref{eq:Lag}), (\ref{eq:mass}) and (\ref{eq:lambda})
and the EOM are obtained by performing the variation. In both cases one gets
\be
\left\{
\begin{array}{l}
\displaystyle
\rule{0pt}{1.5em}D^{\bar\mu}\left(a(z) D_{\bar\mu} \phi\right)
+ \frac{a(z)}{r^2}\phi(1-|\phi|^2) + i\gamma L \epsilon^{\bar\mu \bar\nu}
\partial_{\bar\mu}\left(\frac{s}{r}\right)D_{\bar\nu}\phi = 0\\
\displaystyle
\rule{0pt}{1.5em}\partial^{\bar\mu}\left(r^2 a(z) A_{\bar\mu\bar\nu}\right)
- a(z) \left(i \phi^\dagger D_{\bar\nu} \phi + h.c.\right)
+ \gamma L \epsilon^{\bar\mu \bar\nu} \partial_{\bar\mu}\left(\frac{s}{r}\right)
(|\phi|^2-1)=0\\
\displaystyle
\rule{0pt}{1.5em}\partial_{\bar\mu}\left(a(z)\partial^{\bar\mu}s\right)
-\frac{\gamma L}{2 r}\epsilon^{\bar\mu\bar\nu}\left[
\partial_{\bar\mu}(-i \phi^\dagger D_{\bar\nu}\phi + h.c.) + A_{\bar\mu\bar\nu}\right]
= 0
\end{array}
\right.\,,
\ee
for the fields which are already ``turned on" in the static case.
For the ``new'' fields which appear in the rotating skyrmion solution we have
\bea
\left\{
\begin{array}{l}
\displaystyle
\partial^{\bar\mu}(r^2 a(z) \partial_{\bar\mu} v) - 2 a(z)\left[
v(1+\left|\phi\right|^2) - \chi\phi^\dagger-\phi\chi^\dagger\right]
\\ \displaystyle
\rule{0pt}{1.5em}\hspace{5cm}
+\gamma L \epsilon^{\bar\mu \bar\nu}\left[
\frac{1}{2}(\left|\phi\right|^2 - 1)
B_{\bar\mu \bar\nu}+ r Q
A_{\bar\mu \bar\nu}\right]=0\\
\displaystyle
\rule{0pt}{1.5em}D^{\bar\mu}(r^2 a(z) D_{\bar\mu} \chi) + a(z)\left[2v\phi
-(1+\left|\phi\right|^2)\chi\right]
-\gamma L \epsilon^{\bar\mu \bar\nu}(D_{\bar\mu}\phi)
\left[i \partial_{\bar\nu}(r Q)+ D_{\bar\nu}\rho\right]=0\\
\displaystyle
\rule{0pt}{1.5em}\frac{1}{r}\partial^{\bar\mu}(r^2 a(z) \partial_{\bar\mu} Q)
-\frac{2}{r}a(z) Q-\frac{\gamma L}{2}\epsilon^{\bar\mu \bar\nu}\Big[
(i D_{\bar\mu}\phi (D_{\bar\nu}\chi)^\dagger + h.c.)
\\
\displaystyle
\hspace{11em}
+ \frac{1}{2}A_{\bar\mu \bar\nu}(2 v - \chi\phi^\dagger-\phi\chi^\dagger)
- \frac{2}{\alpha^2}D_{\bar\mu}\rho\, \partial_{\bar\nu}\left(\frac{s}{r}\right)
\Big]=0\\
\displaystyle
\rule{0pt}{1.5em}\partial_{\bar \mu}(a(z) D_{\bar\mu}\rho)
-\frac{\gamma L}{2}\epsilon^{\bar\mu \bar\nu}\Big[
\left(D_{\bar\mu}\phi(D_{\bar\nu}\chi)^\dagger + h.c.\right)
+\frac{i}{2}A_{\bar\mu \bar\nu}(\phi\chi^\dagger - \chi \phi^\dagger)
\\
\displaystyle
\rule{0pt}{1.5em}
\hspace{21em}
+\frac{2}{\alpha^2} \partial_{\bar\mu}(r Q)\partial_{\bar\nu}\left(\frac{s}{r}\right)\Big]=0\\
\displaystyle
\rule{0pt}{1.5em}\partial^{\bar\nu}\left(r^2 a(z)B_{\bar\nu\bar\mu}\right)
+ 2 a(z) D_{\bar\mu} \rho\\
\displaystyle
\hspace{3em}+\gamma L \epsilon^{\bar\mu\bar\nu}
\Big\{\left[(\chi-v \phi)(D_{\bar\nu}\phi)^\dagger + h.c.\right]
+(1-|\phi|^2)\partial_{\bar\nu} v -\frac{2 r}{\alpha^2} Q\, \partial_{\bar\nu}\left(\frac{s}{r}\right)
\Big\} =0
\end{array}
\right.
\eea

In order to solve numerically the EOM, they must be rewritten as
a system of elliptic partial differential equations. This can be achieved by
choosing a 2D Lorentz gauge condition for the residual $U(1)$ gauge fields
\be
\partial^{\bar\mu} A_{\bar\mu}=0\,,
\qquad \quad
\partial^{\bar\mu} B_{\bar\mu}=0\,.
\label{eq:2Dgauge}
\ee
The equations for $A_{\bar\nu}$ become
$J^{\bar\nu} = \partial_{\bar\mu}\left(r^2 a A^{\bar\mu\bar\nu}\right)
= r^2 a \partial_{\bar\mu}\partial^{\bar\mu} A^{\bar\nu}
+\partial_{\bar\mu}(r^2 a) A^{\bar\mu\bar\nu}$ which is an elliptic equation
and a similar result is obtained for $B_{\bar\mu}$.

The gauge condition needs only to be imposed at the boundaries, while 
in the bulk one can just solve the ``gauge-fixed'' EOM treating the two 
gauge field components as independent. The fact that the currents are 
conserved, $\partial_{\bar\nu}J^{\bar\nu}=0$, implies indeed an 
elliptic equation for $\partial^{\bar\mu} A_{\bar\mu}$ which has a 
unique solution once the boundary conditions are specified.
If imposed on the boundary, therefore, the gauge conditions
are maintained also in the bulk.

The IR and UV boundary conditions on the 2D fields follow from
Eq.~(\ref{irboundary condition}) and Eq.~(\ref{uvboundary condition})
and from the gauge choice in Eq.~(\ref{eq:2Dgauge}).
They are given explicitly by
\be
z=z_{\IR}\ :
\quad
\left\{
\begin{array}{l}
\phi_1 = 0\\
\partial_2 \phi_2 = 0\\
A_1 = 0\\
\partial_2 A_2 = 0\\
\partial_2 s = 0
\end{array}
\right.
\qquad\qquad
\left\{
\begin{array}{l}
\chi_1 = 0\\
\partial_2 \chi_2 = 0\\
\partial_2 v = 0\\
\partial_2 Q = 0
\end{array}
\right.
\qquad\qquad
\left\{
\begin{array}{l}
\rho = 0\\
B_1 = 0\\
\partial_2 B_2 = 0
\end{array}
\right.\,,
\label{eq:bcir}
\ee
and
\be
z=z_{\UV}\ :
\quad
\left\{
\begin{array}{l}
\phi_1 = 0\\
\phi_2 = -1\\
A_1 = 0\\
\partial_2 A_2 = 0\\
s=0
\end{array}
\right.
\qquad\qquad
\left\{
\begin{array}{l}
\chi_1 = 0\\
\chi_2 = -1\\
v = -1\\
Q = 0
\end{array}
\right.
\qquad\qquad
\left\{
\begin{array}{l}
\rho = 0\\
B_1 = 0\\
\partial_2 B_2 = 0
\end{array}
\right.\,.
\label{eq:bcuv}
\ee

The boundary conditions at $r=\infty$ have to ensure that the energy of the
solution is finite; this means that the fields should approach a pure-gauge
configuration. At the same time one has to require that the solution is non-trivial
and its topological charge (Eq.~(\ref{Bch})) is equal to one. We have
\be
r=\infty\ :
\quad
\left\{
\begin{array}{l}
\phi = -i e^{i \pi z/L}\\
\partial_1 A_1 = 0\\
A_2 = \frac{\pi}{L}\\
s=0
\end{array}
\right.
\quad\qquad
\left\{
\begin{array}{l}
\chi = i e^{i \pi z/L}\\
v = -1\\
Q = 0
\end{array}
\right.
\qquad\quad
\left\{
\begin{array}{l}
\rho = 0\\
\partial_1 B_1 = 0\\
B_2 = 0
\end{array}
\right.\,.
\label{eq:bcrinfty}
\ee

The $r=0$ boundary of our domain requires an ad hoc treatment, given that
the EOM become singular there. Of course this boundary is
not a true boundary of our 5D space, but it represents some internal points.
Thus we must require the 2D solution
to give rise to regular 5D vector fields at $r=0$ and we must also require
the gauge choice to be fulfilled. These conditions are
\be
r=0\ :
\quad
\left\{
\begin{array}{l}
\phi_1/r \rightarrow A_1\\
(1+\phi_2)/r \rightarrow 0\\
A_2 = 0\\
\partial_1 A_1 = 0\\
s=0
\end{array}
\right.
\quad\qquad
\left\{
\begin{array}{l}
\chi_1 = 0\\
\chi_2 = -v\\
\partial_1 \chi_2 = 0\\
Q = 0
\end{array}
\right.
\qquad\qquad
\left\{
\begin{array}{l}
\rho/r \rightarrow B_1\\
\partial_1 B_1 = 0\\
B_2 = 0
\end{array}
\right.\,.
\label{eq:bcr0}
\ee

\subsection{COMSOL Implementation}

To obtain the numerical solution of the EOM we used the
COMSOL 3.4 package \cite{comsol}, which permits to solve a generic system
of differential elliptic equations by the finite elements method.
A nice feature of this software is that it allows us to extend the
domain up to boundaries where the EOM are singular
({\it i.e.} the $r=0$ line), because it does not use the bulk equations
on the boundaries, but, instead, it imposes the boundary conditions.

In order to improve the convergence of the program and the numerical accuracy,
one is forced to perform a coordinate and a field redefinition. The former
is needed to include the $r=\infty$ boundary in the domain in which the numerical
solution is computed. The advantage of this procedure is the fact that in this way
one can correctly enforce the right behaviour of the fields at infinity by
imposing the $r=\infty$ boundary conditions. A convenient coordinate change is
given by
\be
x = c \arctan\left(\frac{r}{c}\right)\,,
\ee
where $x$ is the new coordinate used in the program and $c$ is an arbitrary constant.
The domain in the $x$ direction is now reduced to the interval $[0,c \pi/2]$.
The parameter $c$ has been introduced to improve the numerical convergence of the
solution. A good choice for $c$ is $c\sim 10$, which allows to have
a reasonable domain for $x$ and, at the same time, does not compress the solution
towards $x=0$.

A field redefinition is needed to impose the regularity conditions at $r=0$
(Eq.~(\ref{eq:bcr0})). For this purpose we use the rescaled fields
\be
\left\{
\begin{array}{l}
\phi_1 = x \psi_1\\
\phi_2 = -1 + x \psi_2\\
\rho = x \tau
\end{array}
\right.\,.
\ee
With these redefinitions, in the new coordinates, the $r=0$ boundary conditions
read as
\be
r=0\ :
\quad
\left\{
\begin{array}{l}
\psi_1 - A_1 = 0\\
\psi_2 = 0\\
A_2 = 0\\
\partial_x A_1 = 0
\end{array}
\right.
\quad\qquad
\left\{
\begin{array}{l}
\chi_1 = 0\\
\partial_x \chi_2 = 0\\
v = -\chi_2\\
Q = 0
\end{array}
\right.
\qquad\qquad
\left\{
\begin{array}{l}
\tau - B_1 = 0\\
\partial_x B_1 = 0\\
B_2 = 0
\end{array}
\right.\,.
\label{eq:bcr0new}
\ee

In order to ensure the convergence of the program another modification is needed.
As already discussed, to obtain a soliton solution with non-vanishing topological
charge we have to impose non-trivial boundary conditions for the 2D fields
at $r=\infty$ (Eq.~(\ref{eq:bcrinfty})). It turns out that if imposing such conditions
the program is not able to reach a regular solution. This is so because the
$r=\infty$ boundary is singular and imposing
non-trivial (though gauge-equivalent to the trivial ones) boundary conditions
at a singular point spoils the regularity of the numerical solution; the same would
happen if the topological twist was located at $r=0$. To fix this problem we
have to perform a gauge transformation which reduces the $r=\infty$ conditions to
trivial ones and preserves the ones at $r=0$
 at the cost of introducing a ``twist'' on the UV boundary.
For this, we use a transformation of the residual $U(1)$ chiral
gauge symmetry associated to $SU(2)_{L,R}$ (Eq.~(\ref{eq:resU1SU2})) with
\be
\alpha(r,z) = (1-z/L) f(r)\,,
\ee
where $f(r)$ can be an arbitrary function which respects the conditions
\be
\left\{
\begin{array}{l}
f(0) = 0\\
f(\infty) \rightarrow \pi
\end{array}
\right.
\qquad\text{and}\qquad
\left\{
\begin{array}{l}
f''(0) = 0\\
f''(\infty) \rightarrow 0
\end{array}
\right.\,.
\ee
For $c\sim 10$ it turns out that a good choice for $f(r)$ is $f(r) = 2 \arctan r$.
The gauge-fixing condition for $A_{\bar\mu}$ is now modified as
\be
\partial_r A_1 + \partial_z A_2 - (1-z/L) f''(r) = 0\,,
\label{eq:newgauge}
\ee
the UV boundary conditions are given by
\be
z=z_{\UV}\ :
\quad
\left\{
\begin{array}{l}
x \psi_1 = \sin f(r)\\
(-1 + \psi_2) = -\cos f(r)\\
A_1 = f'(r)\\
\partial_z A_2 = 0\\
s=0
\end{array}
\right.
\quad\quad
\left\{
\begin{array}{l}
\chi_1 = -\sin f(r)\\
\chi_2 = \cos f(r)\\
v = -1\\
Q = 0
\end{array}
\right.
\quad\quad
\left\{
\begin{array}{l}
\tau = 0\\
B_1 = 0\\
\partial_z B_2 = 0
\end{array}
\right.\,,
\label{eq:bcuvnew}
\ee
and the $r=\infty$ constraints are now trivial
\be
r=\infty\ :
\quad
\left\{
\begin{array}{l}
\psi_1 = 0\\
(-1 + x \psi_2) = 1\\
\partial_x A_1 = 0\\
A_2 = 0\\
s=0
\end{array}
\right.
\qquad\qquad
\left\{
\begin{array}{l}
\chi = -i\\
v = -1\\
Q = 0
\end{array}
\right.
\qquad\qquad
\left\{
\begin{array}{l}
\tau = 0\\
\partial_x B_1 = 0\\
B_2 = 0
\end{array}
\right.\,,
\label{eq:bcrinftynew}
\ee
whereas the $r=0$ and the IR boundary conditions are left unchanged.
Notice that in the new gauge the EOM for $A_{\bar\mu}$ are modified
in accord to Eq.~(\ref{eq:newgauge}), however they are still in the
form of elliptic equations.

%%%%%%%%%%%%%%%


\begin{thebibliography}{99}
  

\bibitem{'tHooft:1973jz}
  G.~'t Hooft,
  %``A PLANAR DIAGRAM THEORY FOR STRONG INTERACTIONS,''
  Nucl.\ Phys.\  B {\bf 72} (1974) 461.
  %%CITATION = NUPHA,B72,461;%%


\bibitem{Adkins:1983ya}
  G.~S.~Adkins, C.~R.~Nappi and E.~Witten,
  %``Static Properties Of Nucleons In The Skyrme Model,''
  Nucl.\ Phys.\  B {\bf 228} (1983) 552.
  %%CITATION = NUPHA,B228,552;%%


\bibitem{Skyrme:1961vq}
  T.~H.~R.~Skyrme,
  %``A Nonlinear field theory,''
  Proc.\ Roy.\ Soc.\ Lond.\  A {\bf 260} (1961) 127.
  %%CITATION = PRSLA,A260,127;%%


\bibitem{Meissner:1987ge}
  For a review see U.~G.~Meissner,
  %``Low-Energy Hadron Physics From Effective Chiral Lagrangians With Vector
  %Mesons,''
  Phys.\ Rept.\  {\bf 161} (1988) 213.
  %%CITATION = PRPLC,161,213;%%


\bibitem{Igarashi:1985et}
    Y.~Igarashi, M.~Johmura, A.~Kobayashi, H.~Otsu, T.~Sato and S.~Sawada,
  %``Stabilization Of Skyrmions Via Rho Mesons,''
  Nucl.\ Phys.\  B {\bf 259} (1985) 721.
  %%CITATION = NUPHA,B259,721;%%

\bibitem{Adkins:1983nw}
  G.~S.~Adkins and C.~R.~Nappi,
  %``Stabilization Of Chiral Solitons Via Vector Mesons,''
  Phys.\ Lett.\  B {\bf 137} (1984) 251.
  %%CITATION = PHLTA,B137,251;%%

\bibitem{Maldacena:1997re}
J.~M.~Maldacena,
%``The large N limit of superconformal field theories and supergravity,''
Adv.\ Theor.\ Math.\ Phys.\  {\bf 2} (1998) 231.
%%CITATION = HEP-TH 9711200;%%



\bibitem{Gubser:1998bc}
S.~S.~Gubser, I.~R.~Klebanov and A.~M.~Polyakov,
%``Gauge theory correlators from non-critical string theory,''
Phys.\ Lett.\ B {\bf 428} (1998) 105.
%%CITATION = HEP-TH 9802109;%%

\bibitem{Witten:1998qj}
  E.~Witten,
  %``Anti-de Sitter space and holography,''
  Adv.\ Theor.\ Math.\ Phys.\  {\bf 2} (1998) 253.
  %%CITATION = 00203,2,253;%%





\bibitem{Erlich:2005qh}
  J.~Erlich, E.~Katz, D.~T.~Son and M.~A.~Stephanov,
  %``QCD and a holographic model of hadrons,''
  Phys.\ Rev.\ Lett.\  {\bf 95} (2005) 261602.
  %%CITATION = PRLTA,95,261602;%%

\bibitem{Rold:2005}
  L.~Da Rold and A.~Pomarol,
  %``Chiral symmetry breaking from five dimensional spaces,''
  Nucl.\ Phys.\  B {\bf 721} (2005) 79; JHEP {\bf 0601} (2006) 157.
  %%CITATION = NUPHA,B721,79;%%
  %%CITATION = JHEPA,0601,157;%%




\bibitem{Pomarol:2007kr}
  A.~Pomarol and A.~Wulzer,
  %``Stable skyrmions from extra dimensions,''
  JHEP {\bf 0803} (2008) 051.
  %%CITATION = JHEPA,0803,051;%%

%\cite{Pomarol:2008aa}
\bibitem{Pomarol:2008aa}
  A.~Pomarol and A.~Wulzer,
  %``Baryon Physics in Holographic QCD,''
  Nucl.\ Phys.\  B {\bf 809} (2009) 347.
%  [arXiv:0807.0316 [hep-ph]].
  %%CITATION = NUPHA,B809,347;%%

%\cite{Panico:2008it}
\bibitem{Panico:2008it}
  G.~Panico and A.~Wulzer,
  %``Nucleon Form Factors from 5D Skyrmions,''
  arXiv:0811.2211 [hep-ph].
  %%CITATION = ARXIV:0811.2211;%%





\bibitem{Hirn:2005nr}
  J.~Hirn and V.~Sanz,
  %``Interpolating between low and high energy QCD via a 5D Yang-Mills  model,''
  JHEP {\bf 0512} (2005) 030.
  %%CITATION = JHEPA,0512,030;%%



\bibitem{Nawa:2006gv}
  K.~Nawa, H.~Suganuma and T.~Kojo,
  %``Baryons in Holographic QCD,''
  Phys.\ Rev.\  D {\bf 75} (2007) 086003;
%  [arXiv:hep-th/0612187].
  %%CITATION = PHRVA,D75,086003;%%
  %``Brane-induced Skyrmions: Baryons in holographic QCD,''
  Prog.\ Theor.\ Phys.\ Suppl.\  {\bf 168} (2007) 231.
  %%CITATION = PTPSA,168,231;%%


\bibitem{Hata:2007mb}
  H.~Hata, T.~Sakai, S.~Sugimoto and S.~Yamato,
  %``Baryons from instantons in holographic QCD,''
  arXiv:hep-th/0701280.
  %%CITATION = HEP-TH/0701280;%%


\bibitem{Hong:2006ta}
  D.~K.~Hong, T.~Inami and H.~U.~Yee,
  %``Baryons in AdS/QCD,''
  Phys.\ Lett.\  B {\bf 646} (2007) 165;
%  [arXiv:hep-ph/0609270].
  %%CITATION = PHLTA,B646,165;%%
  D.~K.~Hong, M.~Rho, H.~U.~Yee and P.~Yi,
  %``Chiral dynamics of baryons from string theory,''
  Phys.\ Rev.\  D {\bf 76} (2007) 061901;
%  [arXiv:hep-th/0701276].
  %%CITATION = PHRVA,D76,061901;%%
  %``Nucleon Form Factors and Hidden Symmetry in Holographic QCD,''
  Phys.\ Rev.\  D {\bf 77} (2008) 014030.
  %%CITATION = PHRVA,D77,014030;%%

\bibitem{Hata:2008xc}
  H.~Hata, M.~Murata and S.~Yamato,
  %``Chiral currents and static properties of nucleons in holographic QCD,''
  arXiv:0803.0180 [hep-th];
  %%CITATION = ARXIV:0803.0180;%%
  K.~Hashimoto, T.~Sakai and S.~Sugimoto,
  %``Holographic Baryons : Static Properties and Form Factors from Gauge/String
  %Duality,''
  arXiv:0806.3122 [hep-th].
  %%CITATION = ARXIV:0806.3122;%%

\bibitem{Sakai:2004cn}
  T.~Sakai and S.~Sugimoto,
  %``Low energy hadron physics in holographic QCD,''
  Prog.\ Theor.\ Phys.\  {\bf 113} (2005) 843.
  %%CITATION = PTPKA,113,843;%%


%\cite{Hill:2006wu}
\bibitem{Hill:2006wu}
  C.~T.~Hill,
  %``Exact equivalence of the D = 4 gauged Wess-Zumino-Witten term and the D  =
  %5 Yang-Mills Chern-Simons term,''
  Phys.\ Rev.\  D {\bf 73} (2006) 126009.
  %%CITATION = PHRVA,D73,126009;%%

\bibitem{Panico:2007qd}
  G.~Panico and A.~Wulzer,
  %``Effective Action and Holography in 5D Gauge Theories,''
  JHEP {\bf 0705} (2007) 060.
  %%CITATION = JHEPA,0705,060;%%

%\cite{Bando:1987br}
\bibitem{Bando:1987br}
  M.~Bando, T.~Kugo and K.~Yamawaki,
  %``Nonlinear Realization and Hidden Local Symmetries,''
  Phys.\ Rept.\  {\bf 164} (1988) 217.
  %%CITATION = PRPLC,164,217;%%

\bibitem{Georgi:1989xy}
  H.~Georgi,
  %``Vector Realization Of Chiral Symmetry,''
  Nucl.\ Phys.\  B {\bf 331} (1990) 311.
  %%CITATION = NUPHA,B331,311;%%

%\cite{Ecker:1989yg}
\bibitem{Ecker:1989yg}
  G.~Ecker, J.~Gasser, H.~Leutwyler, A.~Pich and E.~de Rafael,
  %``Chiral Lagrangians for Massive Spin 1 Fields,''
  Phys.\ Lett.\  B {\bf 223} (1989) 425;
  %%CITATION = PHLTA,B223,425;%%
G.~Ecker, J.~Gasser, A.~Pich and E.~de Rafael,
  %``The Role Of Resonances In Chiral Perturbation Theory,''
  Nucl.\ Phys.\  B {\bf 321} (1989) 311.
  %%CITATION = NUPHA,B321,311;%%

\bibitem{Son:2003et}
  D.~T.~Son and M.~A.~Stephanov,
  %``QCD and dimensional deconstruction,''
  Phys.\ Rev.\  D {\bf 69} (2004) 065020.
  %%CITATION = PHRVA,D69,065020;%%


\bibitem{Witten:1976ck}
  E.~Witten,
  %``Some exact multipseudoparticle solutions of classical Yang-Mills  theory,''
  Phys.\ Rev.\ Lett.\  {\bf 38} (1977) 121.

\bibitem{comsol}
See http://www.comsol.com.

\bibitem{Finkelstein:1968hy}
  D.~Finkelstein and J.~Rubinstein,
  %``Connection between spin, statistics, and kinks,''
  J.\ Math.\ Phys.\  {\bf 9} (1968) 1762.
  %%CITATION = JMAPA,9,1762;%%

\bibitem{Braaten:1986iw}
  E.~Braaten, S.~M.~Tse and C.~Willcox,
  %``Electromagnetic Form-Factors In The Skyrme Model,''
  Phys.\ Rev.\ Lett.\  {\bf 56} (1986) 2008.
  %%CITATION = PRLTA,56,2008;%%

\bibitem{Witten:1979kh}
  E.~Witten,
  %``Baryons In The 1/N Expansion,''
  Nucl.\ Phys.\  B {\bf 160} (1979) 57.
  %%CITATION = NUPHA,B160,57;%%

\bibitem{Manohar:1998xv}
  A.~V.~Manohar,
  %``Large N QCD,''
  arXiv:hep-ph/9802419.
  %%CITATION = HEP-PH/9802419;%%


\bibitem{Witten:1983tx}
  E.~Witten,
  %``Current Algebra, Baryons, And Quark Confinement,''
  Nucl.\ Phys.\  B {\bf 223} (1983) 433.
  %%CITATION = NUPHA,B223,433;%%%

\bibitem{Karl:1984cz}
  G.~Karl and J.~E.~Paton,
  %``Naive Quark Model For An Arbitrary Number Of Colors,''
  Phys.\ Rev.\  D {\bf 30}, 238 (1984).
  %%CITATION = PHRVA,D30,238;%%

\bibitem{Beg:1973sc}
  M.~A.~B.~Beg and A.~Zepeda,
  %``Pion radius and isovector nucleon radii in the limit of small pion mass,''
  Phys.\ Rev.\  D {\bf 6} (1972) 2912.
  %%CITATION = PHRVA,D6,2912;%%

\bibitem{Machleidt:1987hj}
For a review see
  R.~Machleidt, K.~Holinde and C.~Elster,
  %``The Bonn Meson Exchange Model for the Nucleon Nucleon Interaction,''
  Phys.\ Rept.\  {\bf 149} (1987) 1.
  %%CITATION = PRPLC,149,1;%%

%\cite{Adkins:1983hy}
\bibitem{Adkins:1983hy}
  G.~S.~Adkins and C.~R.~Nappi,
  %``The Skyrme Model With Pion Masses,''
  Nucl.\ Phys.\  B {\bf 233}, 109 (1984).
  %%CITATION = NUPHA,B233,109;%%

\bibitem{Amaldi:1972vf}
  E.~Amaldi {\it et al.},
  %``Axial-vector form-factor of the nucleon from a coincidence experiment on
  %electroproduction at threshold,''
  Phys.\ Lett.\  B {\bf 41} (1972) 216.
  %%CITATION = PHLTA,B41,216;%%

\bibitem{DelGuerra:1976uj}
  A.~Del Guerra {\it et al.},
  %``Threshold $\pi^+$ electroproduction at high momentum transfer: a
  %determination of the nucleon axial vector form-factor,''
  Nucl.\ Phys.\  B {\bf 107} (1976) 65.
  %%CITATION = NUPHA,B107,65;%%

%\bibitem{Weinberg:2006rq}
%See,  for example,  E.~J.~Weinberg and P.~Yi,
%  %``Magnetic monopole dynamics, supersymmetry, and duality,''
%  Phys.\ Rept.\  {\bf 438} (2007) 65.
%  %%CITATION = PRPLC,438,65;%%




%\cite{Chu:1996fr}
%\bibitem{Chu:1996fr}
%  We are following the notations of 
%  C.~S.~Chu, P.~M.~Ho and B.~Zumino,
  %``Non-Abelian Anomalies and Effective Actions for a Homogeneous Space
  %$G/H$,''
%  Nucl.\ Phys.\  B {\bf 475} (1996)  484.
  %%CITATION = NUPHA,B475,484;%%

%\bibitem{Witten:1976ck}
%  E.~Witten,
  %``Some exact multipseudoparticle solutions of classical Yang-Mills  theory,''
%  Phys.\ Rev.\ Lett.\  {\bf 38} (1977) 121.








  
  
%\bibitem{Yao:2006px}
 % W.~M.~Yao {\it et al.}  [Particle Data Group],
  %%``Review of particle physics,''
%  J.\ Phys.\ G {\bf 33} (2006) 1.
%  %%CITATION = JPHGB,G33,1;%%
  
  
%  \bibitem{Sakai:2005yt}
%  T.~Sakai and S.~Sugimoto,
%  %``More on a holographic dual of QCD,''
%  Prog.\ Theor.\ Phys.\  {\bf 114} (2006) 1083.
%    %%CITATION = PTPKA,114,1083;%%
  
  
%\bibitem{Kaymakcalan:1983qq}
%  O.~Kaymakcalan, S.~Rajeev and J.~Schechter,
%  %``Nonabelian Anomaly And Vector Meson Decays,''
%  Phys.\ Rev.\  D {\bf 30} (1984) 594.
%  %%CITATION = PHRVA,D30,594;%%
  
%\bibitem{Grigoryan:2008up}
%  H.~R.~Grigoryan and A.~V.~Radyushkin,
%  %``Anomalous Form Factor of the Neutral Pion in Extended AdS/QCD Model with
%  %Chern-Simons Term,''
%  arXiv:0803.1143 [hep-ph].
%  %%CITATION = ARXIV:0803.1143;%%
      
  
\end{thebibliography}
\end{document}